\documentclass[10pt,a4paper]{JHEP3}
\pdfoutput=1
\usepackage{scrbase}
\makeatletter
\AfterPackage*{xcolor}{%
  \renewcommand*{\normalcolor}{\if@atdocument\let \current@color
    \default@color \set@color\fi}%
}
\AfterPackage*{color}{%
  \renewcommand*{\normalcolor}{\if@atdocument\let \current@color
    \default@color \set@color\fi}%
}
\makeatother

\usepackage[american]{babel}              
\usepackage[utf8]{inputenc}     
\usepackage[T1]{fontenc}        
\usepackage{amsmath,amstext,amsfonts,amssymb}
\usepackage{bm,dsfont}
\usepackage{xcolor,graphicx}
\usepackage{booktabs,dcolumn}
\usepackage{cite,tensor,units}
\usepackage[version=3]{mhchem}

\definecolor{mdarkgreen}{HTML}{00aa00}
\definecolor{mpurple}{HTML}{800080}

\newcolumntype{C}{>{$}c<{$}}
\newcolumntype{L}{>{$}l<{$}}
\newcolumntype{R}{>{$}r<{$}}             

\newcolumntype{S}[1]{>{$}c<{$}@{\hspace{#1}}}
\newcolumntype{T}[1]{>{$}l<{$}@{\hspace{#1}}}
\newcolumntype{U}[1]{>{$}r<{$}@{\hspace{#1}}}

\newcommand\figref[1]{\figurename~\ref{#1}}
\newcommand\tabref[1]{\tablename~\ref{#1}}

\newcommand\mnewl{\\[.5\baselineskip]}

\DeclareMathAlphabet{\mathbfit}{T1}{cmbr}{b}{n}

\newcommand\ci{\mathrm i}
\newcommand\dd[2][]{\operatorname{d^{#1}\mathit{#2}}}
\newcommand\e[1][]{\,\mathrm{e}^{#1}\,}

\newcommand\vvec[1]{\mathbfit{#1}}
\newcommand\abs[1]{\left\lvert#1\right\rvert}
\newcommand\scp{\bm\cdot}

\newcommand\inv[1]{#1^{-1}}
\newcommand\cc[1]{#1^{\ast}}

\DeclareMathOperator\im{Im}
\DeclareMathOperator\re{Re}

\newcommand\dev[2][]{\operatorname{\frac{d^{#1}}{d\mathit{#2}^{#1}}}}

\newcommand\bracket[3]{\left\langle#1\middle|\:#2\:\middle|#3\right\rangle}
\newcommand\expect[1]{\left\langle#1\right\rangle}

\title{Towards a Holographic Realization of Homes' Law}

\author{Johanna Erdmenger, Patrick Kerner and Steffen Müller\\
  Max-Planck-Institute for Physics (Werner-Heisenberg-Institut),\\
  Föhringer Ring 6, 80805 Munich, Germany\\\vglue0pt
  E-mail: \email{jke@mppmu.mpg.de,pkerner@mppmu.mpg.de,smueller@mppmu.mpg.de}}

\abstract{
  Gauge/gravity duality has proved to be a very successful tool for
  describing strongly coupled systems in particle physics and heavy ion
  physics. The application of the gauge/gravity duality to 
  quantum matter is a promising candidate to explain questions concerning
  non-zero temperature dynamics and transport coefficients. To a large extent,
  the success of applications of gauge/gravity duality to the quark-gluon
  plasma is founded on the derivation of a universal result, the famous ratio 
  of shear viscosity and entropy density. As a base for applications to
  condensed matter physics, it is highly desirable to have a similar universal
  relation in this context as well. A candidate for such a universal law is
  given by \textit{Homes' law}: High $T_c$ superconductors, as well as some
  conventional superconductors, exhibit a universal scaling relation between
  the superfluid density at zero temperature and the conductivity at the
  critical temperature times the critical temperature itself. In this
  work we describe progress in employing the
  models of holographic superconductors to realize Homes' law and to find a
  universal relation governing strongly correlated quantum matter. We 
  calculate diffusive processes, including the
  backreaction of the gravitational matter fields on the
  geometry. We consider both holographic s-wave and p-wave
  superconductors. We show that a particular form of Homes' law holds
  in the absence of backreaction. Moreover, we suggest further steps
  to be taken for holographically realizing Homes' law more generally 
  in the presence of backreaction.
}

\preprint{MPP-2012-100}

\begin{document}

\section{Introduction}

Gauge/gravity duality has proved to be a valuable tool for exploring
strongly coupled regimes of field theories. The best studied example so far for
applications to experimentally accessible strongly coupled systems is the
application to the quark-gluon plasma. A very important example for this is the derivation of the famous result for the ratio of the shear viscosity and the entropy density \cite{Kovtun:2003wp},
\begin{equation}
  \frac\eta s=\frac1{4\pi}\frac\hbar{k_{\text B}}.
  \label{eq:Eta-Over-S}
\end{equation}
Here the physical constants $\hbar$ and $k_{\text B}$ are written out explicitly
in order to illustrate the influence of quantum mechanics and thermal
physics.\footnote{Subsequently, we will set the physical constants $c$, 
  $\hbar$ and $k_{\text B}$ to one in the following sections except for\\
  Section \ref{ssec:Drude-Sommerfeld} and Section 
  \ref{ssec:Homes-Law-Condensed-Matter}.}
It has been shown \cite{Buchel:2003tz,Kovtun:2004de,Iqbal:2008by} that this
result applies universally for any isotropic gauge/gravity duality model based
on an Einstein-Hilbert action on the gravity side. Exceptions are found by 
considering higher curvature corrections \cite{Sinha:2009ev} or anisotropic
configurations, see\cite{Erdmenger:2010xm,Erdmenger:2011tj} and
\cite{Rebhan:2011vd}. Recently, the focus of applying the tools of the
gauge/gravity duality has been widened to other strongly coupled systems in
physics, especially to problems in condensed matter physics. In particular,
significant progress has been made in describing holographic fermions (see
\cite{Liu:2009dm,Cubrovic:2009ye,Iqbal:2011ae} and references therein),
superconductors/superfluids (for instance \cite{Hartnoll:2009sz,Herzog:2009xv,Horowitz:2010gk,Kaminski:2010zu,She:2011cm}
and references therein) and to some extent also to lattices \cite{Horowitz:2012ky,Liu:2012tr}. For obtaining a solid general framework for
condensed matter applications of the gauge/gravity duality, it would be very
useful to derive a universal relation, similar in importance to
\eqref{eq:Eta-Over-S}, designed in particular for applications in condensed
matter physics. Interestingly, the result \eqref{eq:Eta-Over-S} may be
understood in the context of condensed matter physics by a time scale
argument. Here, the properties of quantum critical regions \cite{Sachdev:2011ME,Sachdev:2011cs} give rise to a universal lower bound
\begin{align}
  \tau_\hbar&=\frac\hbar{k_{\text B}T},
  \label{eq:Tau-Hbar}
  \intertext{sometimes called ``Planckian dissipation'' \cite{Zaanen:2004nat}
    which can be compared to the possible lower bound for $\nicefrac\eta s$ 
    given in \eqref{eq:Eta-Over-S}. This seems to imply that the ``strange
    metal phase'' is a nearly perfect fluid without a quasi-particle 
    description as is the quark-gluon plasma, since both cases do not allow for
    long-lived excitations compared to the energy}
  \frac\hbar\tau&\ll\epsilon,
  \label{eq:Quasiparticle-Condition}
  \intertext{but rather describe a regime characterized by}
  \tau&\sim\frac\hbar\epsilon.
  \label{eq:Quantum-Critical-Regime}
  \intertext{In the case of the quark-gluon plasma, a possible characteristic 
    time scale can be defined by}
  \eta&\sim\epsilon\tau.
  \label{eq:Tau-Eta}
\end{align}
In typical condensed matter problems at quantum critical points, the relevant
energy scale $\epsilon$ is set by the thermal energy $\epsilon\sim k_{\text B}T$.
An interesting, yet unresolved problem, are the high temperature 
superconductors and their possible relation to quantum critical regions. A very interesting universality shown by almost all types of superconductors
is {\it Homes' law}. As explained in Section \ref{sec:Homes-Law} in 
detail, this shows some connections to quantum critical regions and the
``Planckian dissipation'' time $\tau_\hbar$. Thus, it is an exciting candidate
to find a universal relation for strongly coupled condensed matter systems
\cite{Ammon:2010zz} where the usual quasi-particle picture seems to fail.
\\[\baselineskip]
In this paper, we outline how Homes' law may be implemented in holography. We
follow a simple approach which allows to demonstrate the validity of Homes' law
in the absence of backreaction of the matter fields on the geometry. We also
find that our straightforward approach requires modifications in the presence 
of backreaction. We present and discuss possible generalizations and indicate
directions for further research. The paper is organized as follows. In Section
\ref{sec:Homes-Law} we give a self-contained exposition of Homes' law and
discuss related condensed matter concepts and their holographic realization. In
the Sections \ref{sec:S-Wave} and \ref{sec:P-Wave} we consider holographic
s-wave and p-wave superconductors, respectively. In particular, we focus on the
effects that arise from the backreaction of the gravitational interaction with
the matter fields onto the geometry, governing the gravity dual. For this
purpose, we numerically determine the phase diagram where we use the strength 
of the backreaction as parameter in addition to the ratio of temperature and
chemical potential. This sets the ground for the calculations of various
diffusion constants, which are discussed for the s-wave superconductor in
Section \ref{ssec:Discussion} and for the p-wave superconductor in Section
\ref{ssec:Diffusion-Constants}. We focus on the critical diffusion and
associated time scales, i.e.~the diffusion at the critical ratio of temperature
and chemical potential depending on the backreaction at which the system
transits to the superconducting phase. We find that a particular version of
Homes' law is satisfied if the backreaction is absent, while further work is
required for the case with backreaction, as we explain. Finally, in Section
\ref{sec:Conclusions} we summarize our results and give some possible
explanations why Homes' law is not confirmed in the approach considered here
once the backreaction on the geometry is turned on. Some extensions of our
original setup to remedy this are discussed in Section \ref{sec:Outlook} along
with some new perspectives on holographic superfluids/superconductors that 
might be of interest to pursue on their own.

%%% Local Variables: 
%%% mode: latex
%%% TeX-master: "homes-law"
%%% End: 

\section{Homes' Law\label{sec:Homes-Law}}

With the discovery of high temperature superconductors, the new era heralded by
the discovery of novel phases of (quantum) matter boosted the need for a new
understanding of the classification of condensed matter systems. The
experimental progress in controlling strongly correlated electronic systems and
the exploration of strongly coupled fermionic/bosonic systems with the help of
ultracold gases presented a new picture of nature that shook the old 
foundations of traditional condensed matter theory, i.e.~Landau's Fermi liquid
theory and transitions between different phases classified by their symmetries.
Famous examples departing from this old scheme are -- apart form the high
temperature superconductors -- topological insulators, quantum critical regions
connected to quantum critical points, and (fractional) quantum Hall effects,
just to name a few (see \cite{Wen:2004ym,Hasan:2010rmp,Sachdev:2011ME,
  Sachdev:2011cs,Zaanen:2011me} and references therein). As in particle 
physics, modern condensed matter theory is concerned with  low-energy
excitations that are described most efficiently by quantum field theories which
reveal the universality of very different microscopic quantum many-body
systems. However, in the strongly interacting cases, the ``mapping'' between 
the relevant degrees of freedom in the low-energy regime and the microscopic
degrees of freedom are far from being clear and understood. Furthermore, it
seems that quantum field theory alone is not enough to tackle strongly
correlated systems and to explain these new states of (quantum)
matter. Interestingly, the universality of Homes' law seems to go beyond the
artificial distinction between traditional and modern condensed matter physics
since it displays a relation that works for conventional superconductors and
high temperature superconductors which can be regarded as representatives of 
the old and the yet to be developed framework.

\subsection{Homes' Law in Condensed Matter
  \label{ssec:Homes-Law-Condensed-Matter}}
An interesting phenomenon to look for universal behavior in condensed matter 
systems, as advertised in the introduction to this section, is the universal
scaling law for superconductors empirically found by Homes et
al.~\cite{Homes:2004nat} by collecting experimental results. This so-called \textit{Homes' law} describes a relation between different quantities of
conventional and unconventional superconductors, i.e.~the superfluid density
$\rho_{\text s}$ at zero temperature and the conductivity $\sigma_{\text{DC}}$
times the critical temperature $T_c$,
\begin{equation}
  \rho_{\text s}=C\sigma_{\text{DC}}(T_c)T_c,
  \label{eq:Homes-Law}
\end{equation}
where Homes et al.~report two different values of the constant $C$ in units
$\unit{[cm]^{-2}}$ for the different cases considered in \cite{Homes:2005prb}.
The value $C=35$ is true for in-plane cuprates and elemental BCS
superconductors, whereas for the cuprates along the c-axis and the dirty limit
BCS superconductors they find $C=65$. The superfluid density $\rho_{\text s}$ is
a measure for the number of particles contributing to the superfluid phase. It
can be thought of as the square of the plasma frequency\footnote{The 
  definition of the plasma frequency and its relations to the dielectric
  function and superconductivity is discussed in detail in Appendix 
  \ref{sec:Plasma-Frequency}.} of the superconducting phase
$\omega_{\text{Ps}}^2$, because the superconductor becomes ``transparent'' for
electromagnetic waves with frequencies larger than
\begin{equation}
  \rho_{\text s}\equiv\omega_{\text{Ps}}^2=\frac{4\pi n_{\text s}e^2}{m^*}.
  \label{eq:Superconducting-Plasmafrequency}
\end{equation}
Here $n_{\text s}$ denotes the superconducting charge carrier density which
describes the number of superconducting charge carriers per volume (and is very
different from the superfluid density $\rho_{\text s}$), $e$ is the elementary
charge and $m^*$ is the effective mass of the charge carrier renormalized due to
interactions. Another way to think about the superfluid density $\rho_{\text s}$
is the London penetration depth $\lambda_{\text L}$ which is basically the
inverse of the superconducting plasma frequency, such that frequencies larger
than $\omega_{\text{Ps}}$ correspond to length scales smaller than
$\lambda_{\text L}$, i.e.~$\rho_{\text s}\equiv\lambda_{\text L}^{-2}$. The
critical temperature $T_c$ is determined by the onset of superconductivity. The
conductivity $\sigma_{\text{DC}}$ and the superconducting plasma frequency
$\omega_{\text{Ps}}$ are for instance obtained from reflectance measurements by
extrapolation to the $\omega\to0$ limit of the complex optical conductivity
$\sigma(\omega)$,
\begin{align}
  \sigma_{\text{DC}}&=\lim_{\omega\to0}\re\sigma(\omega), &
  \omega_{\text{Ps}}^2&=\lim_{\omega\to0}\big(-\omega^2\re\epsilon(\omega)\big),
  \label{eq:Definitions-Sigma-Omega}
\end{align} 
since the high frequency limit of the real part of the dielectric function
$\epsilon(\omega)$ is given by\footnote{A clear introduction to linear
  response, sum rules and the Kramers-Kronig relations\\ in condensed matter 
  can be found in \cite{Altland:2006si}.}
\begin{equation}
  \re\epsilon(\omega)=\epsilon_\infty-\frac{\omega_{\text{Ps}}^2}{\omega^2}.
  \label{eq:Real-Part-Dielectric-Function}
\end{equation}
where $\epsilon_\infty$ is set by the screening due to interband transitions.
\\[\baselineskip]
\begin{figure}[t]
  \centering
  \includegraphics[width=.5\textwidth]{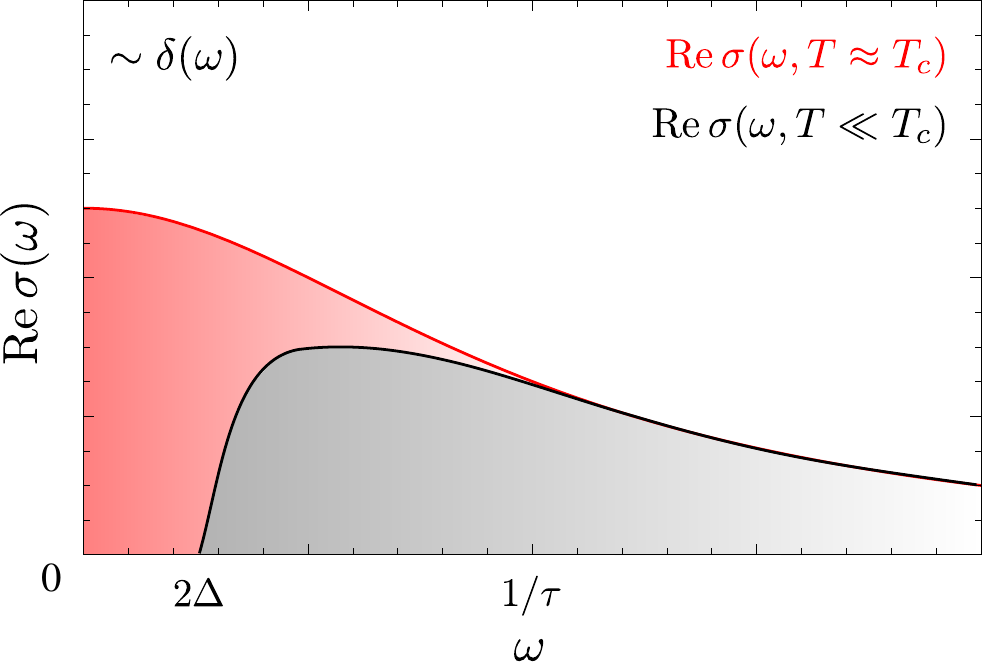}
  \caption{Schematic plots of the optical conductivity above the critical
    temperature $T_c$ and near the absolute zero $T=0$ for a dirty BCS
    superconductor. In the superconducting phase a gap develops for    
    frequencies $\omega<2\Delta$. Note that in the dirty limit the
    quasi-particle scattering rate $\nicefrac1\tau$ is larger than $2\Delta$.
    The missing area (red shaded region) i.e.~the difference between the area 
    under the curve of {\color{red}$\re\sigma(\omega,T\approx T_c)$} and
    $\re\sigma(\omega,T\ll T_c)$ which condenses into the $\delta$-peak at
    $\omega=0$ and thus the superfluid strength (being the coefficient of the 
    $\delta(\omega)$-function) is proportional to the missing area. Following
    the definitions given in \eqref{eq:Definition-Spectral-Weight} we see that
    the area under the $\re\sigma(\omega,T\ll T_c)$ curve determines
    $N_{\text s}$ whereas the red shaded area yields $N_{\text n}-N_{\text s}$.
    \label{fig:Ferrell-Glover-Tinkham}}
\end{figure}
Alternatively, the superconducting plasma frequency may be obtained from the
optical conductivity measured above and below the critical temperature with the
help of the oscillator strength sum rule
\begin{equation}
  \frac{\omega_{\text P}^2}8=\int_0^\infty\dd\omega\re\sigma(\omega),
  \label{eq:Sum-Rule}
\end{equation}
for the optical conductivity. This gives rise to an alternative definition of
the superfluid density as compared to \eqref{eq:Definitions-Sigma-Omega}. We
define the spectral weight in the normal and superconducting phase as follows,
\begin{align}
  N_{\text n}&=\left.\int_0^\infty\dd\omega\re\sigma(\omega)\right|_{T>T_c}
            =\frac{\omega_{\text{Pn}}^2}8, &
  N_{\text s}&=\left.\int_{0^+}^\infty\dd\omega\re\sigma(\omega)\right|_{T<T_c}.
  \label{eq:Definition-Spectral-Weight}
\end{align}
The superfluid density $\rho_{\text s}$ describes the degrees of freedom in the
superconducting phase which have condensed into a Dirac $\delta$-peak at zero
frequency, where  $\rho_{\text s}$ can be viewed as the coefficient of $\delta(\omega)$. This $\delta$-peak in the real part of the conductivity 
gives rise to an infinite DC conductivity or zero resistivity. The
superfluid density is equal to the difference between the integral over the
optical conductivity \eqref{eq:Sum-Rule} evaluated for $T<T_c$ and $T>T_c$ and
generally yields identical values as compared to \eqref{eq:Definitions-Sigma-Omega}. Using the definitions of the spectral weight \eqref{eq:Definition-Spectral-Weight} we find
\begin{equation}
  \rho_{\text s}=8\left(N_{\text n}-N_{\text s}\right).
  \label{eq:Superconducting-Strength}
\end{equation}
This is the \textit{Ferrell-Glover-Tinkham sum rule}. Note that in the
definition of $N_{\text s}$ we have excluded the $\delta$-peak at $\omega=0$
because the oscillator strength sum rule \eqref{eq:Sum-Rule} requires that the area under the optical conductivity curve is identical above
and below $T_c$, i.e.~in the superconducting and normal state. Thus
\eqref{eq:Superconducting-Strength} determines the missing area of the spectral
weight that condensed into the $\delta$-peak at $\omega=0$, as illustrated in
\figref{fig:Ferrell-Glover-Tinkham}. Although the gap describes the creation of
Cooper pairs, it is really the missing area which gives rise to
superconductivity, since according to \eqref{eq:Superconducting-Strength} the
missing area is equal to the degrees of freedom which condense at zero
frequency, thus forming a new (coherent) macroscopic ground state with
off-diagonal long range order. Semiconductors for instance are systems
exhibiting an energy gap in their spectrum as well, but are not necessarily
superconducting since there is no missing area and hence no new ground state
let alone a phase transition. On the other hand, superconductors retain their
properties even if the energy gap is removed (e.g.~by magnetic impurities) due
to the missing area under the $\re\sigma(\omega)$ curve. As a caveat, let us
note that high temperature superconductors may not satisfy the
Ferrell-Glover-Tinkham sum rule, while it is expected to hold for dirty BCS
superconductors\footnote{Experimentally it is not possible to ``integrate'' up
  to $\omega\to\infty$ since a measurement cannot be done at arbitrary high
  frequencies, so in reality we need to introduce a cut-off frequency 
  $\omega_c$. For high temperature superconductors this cut-off frequency may 
  be higher than the experimentally accessible frequencies and thus these 
  superconductors may not satisfy the Ferrell-Glover-Tinkham sum rule
  (see also \cite{Homes:2005prb}).}.
\\[\baselineskip]
In order to see clearly the relation between superfluid density and the product
of the conductivity at the critical temperature and the critical temperature
expressed by Homes' law in \eqref{eq:Homes-Law}, we have reproduced Table I 
from \cite{Homes:2005prb} with and without the elemental superconductors 
niobium \ce{Nb} and lead \ce{Pb} in \figref{fig:Homes-Law-Experimental-Data}.
\begin{figure}[t]
  \centerline{
    \includegraphics[width=.48\textwidth]{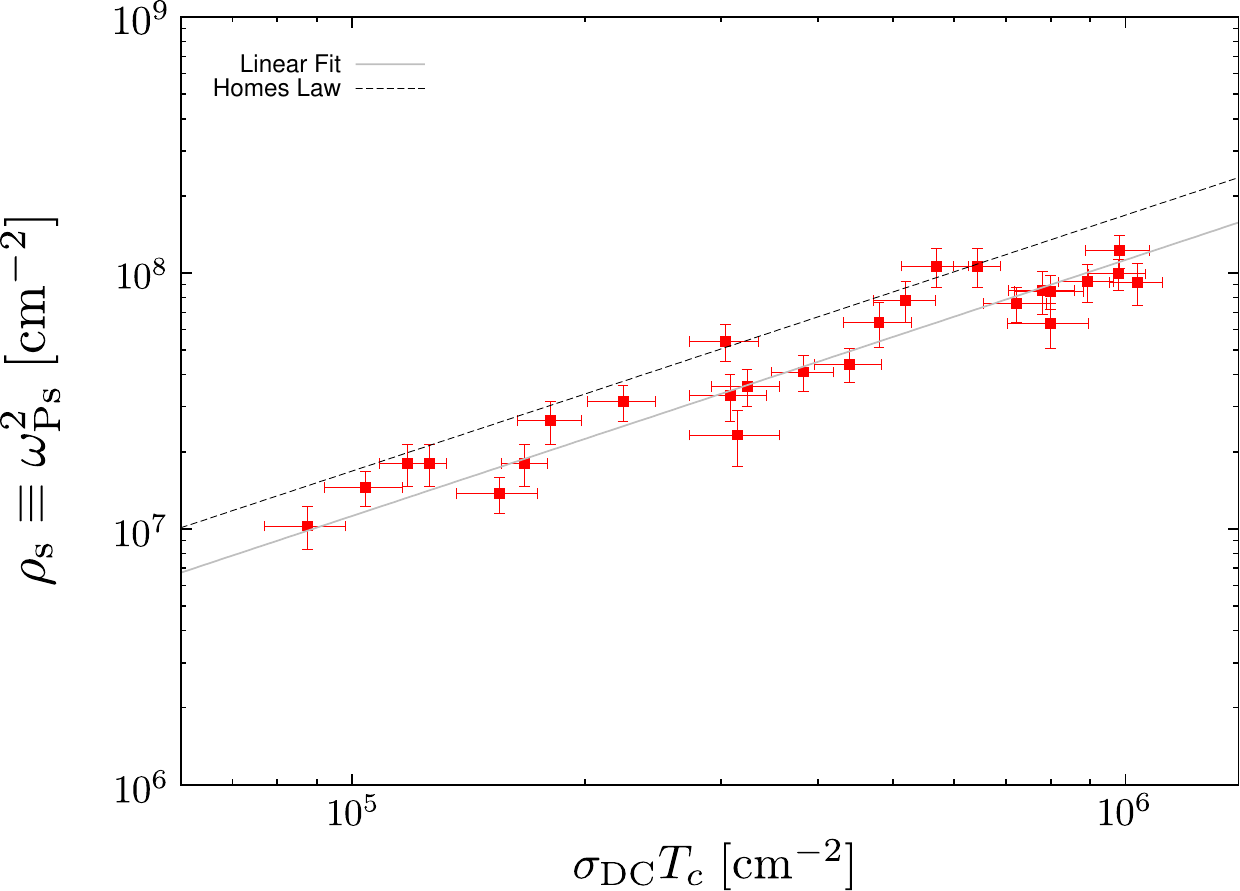}
    \hspace{.04\textwidth}
    \includegraphics[width=.48\textwidth]{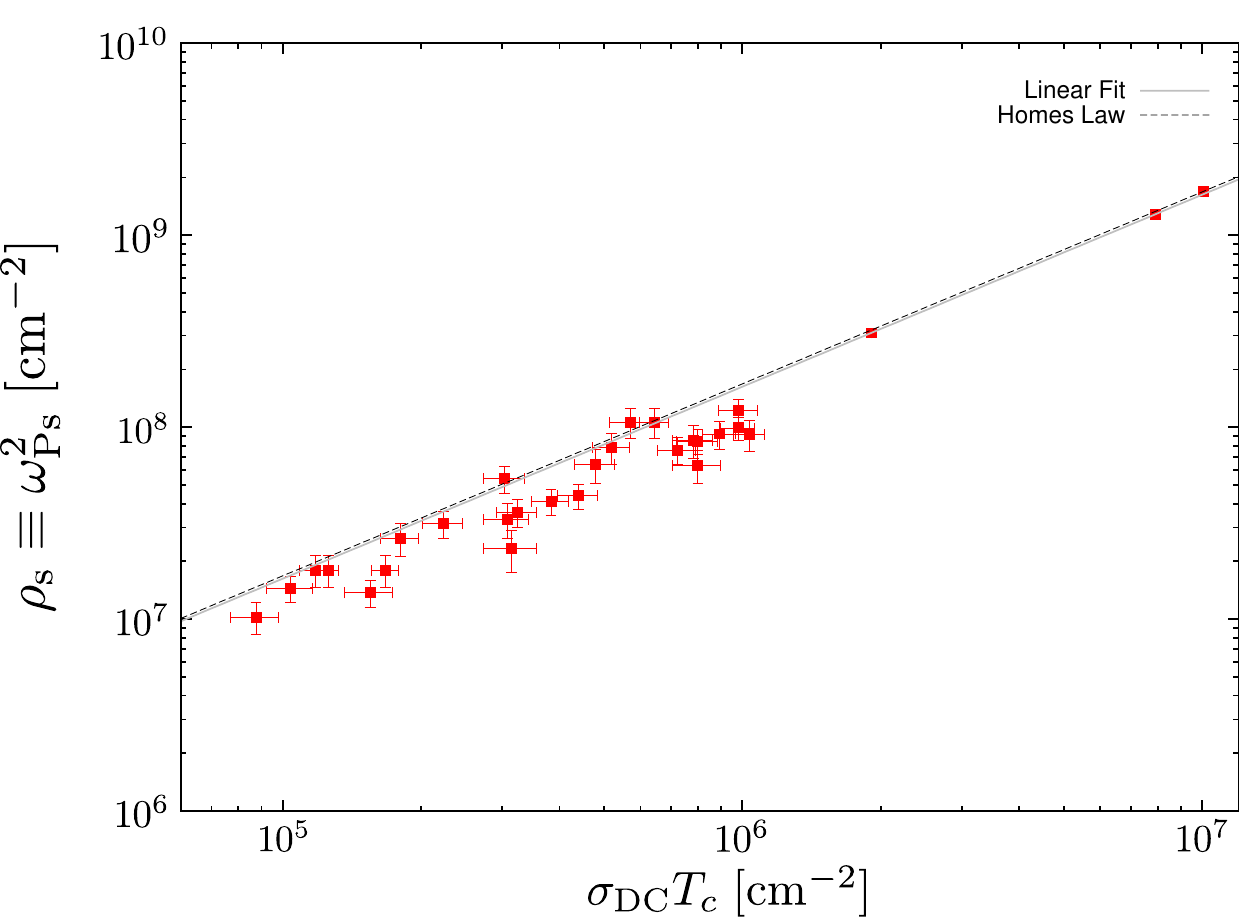}
  }
  \caption{Plots of the complete data in Table 1 in \cite{Homes:2005prb}.    
    The error bars are calculated by 
    $\Delta(\omega_{\text P}^2)=2\omega_{\text P}\Delta\omega_{\text P}$ and 
    $\Delta(\sigma_{\text{DC}}T_c)=T_c\Delta\sigma_{\text DC}$. The left plot
    shows data from high $T_c$ superconductors and for \ce{Ba_{1-x}K_xBiO3}, 
    while the one on the right includes three data points from elemental
    superconductors. As shown on the right the elemental conventional
    superconductors, two data points for \ce{Nb} and one coming from the 
    \ce{Pb} superconductor, actually give Homes' law as stated in 
    \cite{Homes:2005prb}. If these three data points are ignored, as shown in
    the left panel, the linear fit is shifted and the data points are below the
    Homes' law line.\label{fig:Homes-Law-Experimental-Data}}
\end{figure}
Furthermore, in \cite{Homes:2004nat,Zaanen:2004nat,Homes:2005prb,
  Tallon:2006prb} some possible explanations are given concerning the origin of
Homes' law: Conventional dirty-limit superconductors, marginal Fermi-liquid
behavior, for cuprates a Josephson coupling along the c-axis or unitary-limit
impurity scattering. Furthermore, the authors discuss limits where the relation
breaks down, which is true in the overdoped region of cuprates. For dirty limit
BCS superconductors, Homes' law can be explained by the very broad Drude-peak\footnote{The Drude peak is located at
  zero frequency where the real part of $\sigma(\omega)$ reaches its global
  maximum, see for example \figref{fig:Ferrell-Glover-Tinkham} (and for more
  details on the Drude model see Appendix \ref{sec:Drude-Model}).} which is condensing into the superconducting $\delta$-peak at $\omega=0$. The spectral
weight of the condensate may then be estimated by an approximate rectangle of
area $\rho_{\text s}\approx\sigma_{\text{DC}}\cdot2\Delta$ in an optical
conductivity plot, similar to \figref{fig:Ferrell-Glover-Tinkham}, where the gap in the energy of the superconducting state is denoted by $2\Delta$ and
$\sigma_{\text{DC}}$ is the maximum of the curve at $\omega=0$. According to the
BCS model, the energy gap in the superconducting phase is proportional to the
critical temperature $T_c$ and thus $\rho_{\text s}\propto\sigma_{\text{DC}}T_c$. 
For high $T_c$ temperature superconductors the most striking argument can be
found in \cite{Zaanen:2004nat} which links the universal behavior to the
``Planckian dissipation'' giving rise to a perfect fluid description of the
``strange metal phase'' with possible universal behavior, comparable to the
viscosity of the quark-gluon plasma. The argument, reproduced here for
completeness, relies on the fact that the right structure of $\rho_{\text s}$,
$\sigma_{\text{DC}}$ and $T_c$ may be worked out by dimensional analysis: First,
as already stated above \eqref{eq:Superconducting-Plasmafrequency} the
superfluid density must be proportional to the density of the charge carriers 
in the superconducting state. The natural dimension for this quantity is
$(\text{time})^{-2}$ so the product of $\sigma_{\text{DC}}$ and $T_c$ should
have the same physical dimension. Second, the normal state possesses two
relevant time scales, the normal state plasma frequency $\omega_{\text{Pn}}$ and
the relaxation time scale $\tau$, which describes the dissipation of internal
energy into entropy by inelastic scattering. One of the simplest combinations 
is the product of the two time scales which will yield the dimension
$\inv{(\text{time})}$. Therefore we may take the optical conductivity to be of
the Drude-Sommerfeld form (see Appendix \ref{sec:Drude-Model} for details) given by
\begin{equation}
  \sigma_{\text DC}=\frac{\omega_{\text{Pn}}^2\tau}{4\pi}
  =\frac{n_{\text n}e^2\tau}{m^*}.
  \label{eq:Drude-Sommerfeld-Sigma}
\end{equation}
The last and most crucial step is to convert the critical temperature into the
dimension $\inv{(\text{time})}$. Energy and time are related by Heisenberg's
uncertainty principle and thus quantum physics and the idea of ``Planckian
dissipation'' will enter,
\begin{align}
  \inv{\tau_\hbar(T_c)}&=\frac{k_{\text B}T_c}\hbar.
  \label{eq:Planckian-Dissipation}
  \intertext{This time scale is the lowest possible dissipation time for a 
    given temperature. For smaller time scales the system will only allow for
    quantum mechanical dissipationless motion. Interestingly, at finite 
    temperature the lower bound can only be reached if the system is in a 
    quantum critical state \cite{Sachdev:2011ME}. This implies that high $T_c$ 
    superconductors exhibit a quantum critical region above the 
    superconducting dome, which is supported by experimental evidence 
    \cite{Marel:2003nat,Sachdev:2011cs}. To connect the expression on the
    right-hand-side of \eqref{eq:Homes-Law} with the left-hand-side, we can
    employ Tanner's law \cite{Tanner:1998phys} which states that}
  n_{\text s}&\approx\frac14 n_{\text n} \, , 
  \label{eq:Tanners-Law-CMT}
\end{align}
relating the superfluid density to the normal state plasma frequency, as can be
seen from \eqref{eq:Superconducting-Plasmafrequency} and
\eqref{eq:Drude-Sommerfeld-Sigma}. This ``explanation'' of Homes' law will 
guide us in order to find a holographic realization. We will expand on this idea in Section \ref{ssec:Holographic-Homes-Law}.

\subsection{Homes' Law in Holography
  \label{ssec:Holographic-Homes-Law}}
An obstacle towards checking Homes' law directly within holography is that due
to the conformal symmetry and the absence of a lattice, the Drude peak of the
conductivity is given by a delta distribution even at finite temperatures above
the critical temperature, i.e.
\begin{alignat}{3}
  \re\sigma(\omega)&\sim\delta(\omega) &\qquad &\Leftrightarrow &\qquad
  \im\sigma(\omega)&\sim\frac1\omega.
  \label{eq:Drude-Peak}
\end{alignat}
Thus, it is not possible to evaluate $\rho_{\text s}$ directly, which
is related only to the superconducting degrees of freedom condensing at
$\omega=0$. We therefore rewrite Homes' law in such a way that it becomes
accessible to simple models of holography. As we now discuss, this can be
achieved by using the idea of Planckian dissipation following
\cite{Zaanen:2004nat} for high temperature superconductors, as outlined above 
at the end of Section \ref{ssec:Homes-Law-Condensed-Matter}, or by employing 
the Ferrell-Glover-Tinkham sum rule \eqref{eq:Superconducting-Strength} for
dirty BCS superconductors as can be found in \cite{Tallon:2006prb}.
\\[\baselineskip]
For simplicity, let us make two assumptions about holographic superconductors:
First, let us assume that they satisfy the sum rule which requires that the 
area under the optical conductivity curve is identical in the superconducting
phase and the normal phase. Second, we assume that all degrees of freedom
condense in the superconducting state. Using the definition of the spectral
weight in the normal phase $N_{\text n}$ \eqref{eq:Definition-Spectral-Weight}
and the definition of the superconducting plasma frequency
\eqref{eq:Superconducting-Plasmafrequency}, we see that both plasma frequencies
must be equal
\begin{align}
  \omega_{\text{Ps}}^2&=\omega_{\text{Pn}}^2,
  \label{eq:Sum-Rule-Plasma-Frequency}
  \intertext{which implies that $N_{\text s}=0$ in 
    \eqref{eq:Superconducting-Strength}. Here we clearly neglect possible
    missing spectral weight as explained in the second paragraph of Section
    \ref{ssec:Homes-Law-Condensed-Matter}. In the holographic context,
    similar sum rules have been investigated in \cite{Mas:2010ug,
      Gulotta:2010cu,Gursoy:2011gz}. Alternatively, we may assume that the
    holographic superconductors obey Tanner's law \eqref{eq:Tanners-Law}} 
  n_{\text s}&=Bn_{\text n}, 
  \label{eq:Tanners-Law}
  \intertext{where $n_{\text s}$ and $n_{\text n}$ denotes the charge carrier
    density in the superconducting and the normal phase, respectively, and $B$
    is a numerical constant. With either one of these two assumptions,
    we may rewrite Homes' law in such a way that it becomes accessible to
    holography. Let us begin with the assumption that holographic 
    superconductors fulfill the sum rule and that all degrees of freedom are
    participating in the superconducting phase: Starting from}
  \rho_{\text s}\equiv\omega_{\text{Ps}}^2&=C\sigma_{\text{DC}}(T_c)T_c,
  \label{eq:Homes-Law-Sum-Rule}
  \intertext{and assuming that the Drude-Sommerfeld model is still a useful
    approximation (see Appendix \ref{sec:Drude-Model} for details) we can
    replace the conductivity by a characteristic time scale and the plasma
    frequency in the normal conducting phase}
  \sigma_{\text DC}&=\frac{\omega_{\text{Pn}}^2\tau}{4\pi},
  \label{eq:Plasma-Direct-Current}
  \intertext{and thus \eqref{eq:Homes-Law-Sum-Rule} reads}
  \omega_{\text{Ps}}^2&=C\frac{\omega_{\text{Pn}}^2}{4\pi}\tau_cT_c,
  \label{eq:Homes-Law-Sum-Rule2}
  \intertext{where $\tau_c$ denotes $\tau(T_c)$. \eqref{eq:Homes-Law-Sum-Rule2} 
    can be simplified using the assumption 
    \eqref{eq:Sum-Rule-Plasma-Frequency}, }
   \tau_cT_c&=\frac{4\pi}C.
  \label{eq:Homes-Law-Sum-Rule3}
  \intertext{\eqref{eq:Homes-Law-Sum-Rule3} holds if there is no missing
    spectral weight and that all the spectral weight associated with the charge
    carriers condenses in the superconducting phase and hence contributes to 
    the $\delta$-peak. This enables us to identify the plasma frequencies in 
    the two different phases. Alternatively, we can use the assumption that the
    charge carrier densities in both states are proportional to each
    other. Starting again with}
 \rho_{\text s}\equiv4\pi\frac{n_{\text s}e^2}{m^*}&=C\sigma_{\text{DC}}(T_c)T_c,
  \label{eq:Homes-Law-Tanner}
  \intertext{and inserting the Drude-Sommerfeld optical conductivity in terms 
    of the charge carrier density $n_{\text n}$}
  \sigma_{\text DC}&=\frac{n_{\text n}e^2}{m^*}\tau,
  \label{eq:Density-Direct-Current}
  \intertext{we may write \eqref{eq:Homes-Law-Tanner} as}
  n_{\text s}&=\frac C{4\pi}n_{\text n}\tau_cT_c.
  \label{eq:Homes-Law-Tanner2}
  \intertext{Assuming that the holographic superconductors obey
    \eqref{eq:Tanners-Law} we arrive at}
  \tau_cT_c&=\frac{4\pi B}{C}.
  \label{eq:Homes-Law-Tanner3}
  \intertext{Note that the proportionality constants in
    \eqref{eq:Homes-Law-Sum-Rule3} and in \eqref{eq:Homes-Law-Tanner3}
    may not coincide. If they do not, this will indicate that holographic
    superconductors behave either more like in--plane high temperature
    superconductors {or} like dirty-limit BCS superconductors. In any case the
    above simplifications allow us to circumvent the need to calculate spectral
    weights in the superconducting phase or the plasma frequency in either 
    phase (some obstructions are discussed in Section
    \ref{ssec:Drude-Sommerfeld}), and to perform the calculation solely  in 
    the normal phase. Therefore we will extract the time scales in the  normal
    phase of the s- and p-wave superconductors from diffusion constants, 
    basically the momentum and R-charge diffusion denoted by $D_{\text M}$ and 
    $D_{\text R}$, respectively. In particular, since $D(T_c)\propto\tau_c$ we 
    obtain}
  D(T_c)T_c&=\text{const}.
  \label{eq:Homes-Law-Diffusion}
\end{align}
This relation is directly accessible to holography. In fact, without including
the backreaction of the gauge and matter fields on gravity, the diffusion
constants are given by \cite{Kovtun:2003wp,Son:2006em,Son:2007vk}
\begin{align}
  D_{\text M}&=\frac1{4\pi T},  & D_{\text R}&=\frac1{4\pi T}\frac
  d{d-2},
  \label{eq:Diffusion-Constants-Without-Backreaction}
  \intertext{such that}
  D_{\text M}T&=\frac1{4\pi}=\text{const.}, &  
  D_{\text R}T&=\frac1{4\pi}\frac d{d-2}=\text{const.},
  \label{eq:Homes-Law-Diffusion-Without-Backreaction}
\end{align}
where $d$ denotes the dimensionality of the spacetime, and thus Homes' law is
trivially satisfied in this case. This can be derived simply by dimensional
analysis as is done in \cite{Kovtun:2008kx}. Extending the calculation of the
diffusion constants to include backreaction we checked analytically and
numerically that our results reduce to the known results in the limit where the
backreaction vanishes. (see Sections \ref{ssec:Discussion} and
\ref{ssec:Diffusion-Constants} for details).

\subsection{The Drude-Sommerfeld Model and Holography
  \label{ssec:Drude-Sommerfeld}}

The Drude-Sommerfeld model describes the properties of metals (i.a.~electrical/thermal conductivities, heat capacities) by a simple model
incorporating the behavior of the underlying charge carriers. The optical
conductivity of the Drude-Sommerfeld model \eqref{eq:Drude-Sommerfeld-Sigma}
\begin{equation}
  \sigma_{\text DC}=\frac{\omega_{\text{Pn}}^2\tau}{4\pi}
  =\frac{n_{\text n}e^2\tau}{m^*}.
  \label{eq:Drude-Sommerfeld-Sigma2}
\end{equation}
possesses two scales, the plasma frequency $\omega_{\text{Pn}}^2$ setting the
scale above which electromagnetic waves can enter the metal and the relaxation
time scale $\tau$, connected to the mean free path $l=v_{\text F}\tau$ of the
charge carriers (where the charge carrier density is denoted by
$n$). Interactions between the charge carriers are included by the renormalization procedure generating effective masses $m^*$ and correction to
the relaxation time. The inverse of the real part of the optical conductivity
describes absorption processes/loss of energy which is related to dissipation,
whereas the imaginary part describes dispersive processes/change of phase which
is related to energy transport. The resistivity is defined as
$(\re\sigma(\omega=0))^{-1}$ and is the inverse of the maximum of the real part,
the so called Drude peak. The width of the Drude peak is set by the relaxation
rate $\tau^{-1}$. The imaginary part of \eqref{eq:Drude-Sommerfeld-Sigma2} is a
Lorentzian-shaped curve with maximal energy transport at resonance for
$\omega=\tau^{-1}$. For very high frequencies $\omega>\omega_{\text{Pn}}$ the
charge carriers are not able to follow the external excitation and thus the
optical conductivity must approach zero.
\\[\baselineskip]
Comparing our holographic setup to the above condensed matter discussion, we first notice that we do not have an underlying lattice. This implies that the
Drude peak turns into a delta distribution (or $\delta$-peak) at $\omega=0$
since momentum conservation does not allow any dissipative process. In the 
large frequency limit, the optical conductivity approaches the non-vanishing
conformal result, i.e.~$\sigma(\omega\to\infty)=\text{const.}$, since the
conformal symmetry does not permit any scale. In this sense there is no
well-defined plasma frequency above which the system will become transparent,
but a finite absorption due to the non-vanishing real part of the optical
conductivity will be retained. A possible workaround is to define a regulated
plasma frequency \cite{Mas:2010ug} which obey the Kramers-Kronig relations and
thus the sum-rules. Moreover, we cannot compute effective masses $m^*$ or more
generally, we cannot describe any physical parameter related to the
lattice. Additionally, interesting physics is hidden in the zero frequency 
limit of the optical conductivity, which cannot be resolved without a lattice
due to the $\delta$-peak arising from momentum
conservation\footnote{Technically, momentum conservation can be broken by
  neglecting the effects of backreaction onto the geometry on the gravity
  side. Physically this is not very helpful since the fixed geometry introduces
  an artificial dissipative reservoir.}. One way to remedy
these problems is to introduce a lattice explicitly as is done in different 
ways in \cite{Hartnoll:2012rj,Horowitz:2012ky,Liu:2012tr,Iizuka:2012dk}. One of
the results in \cite{Horowitz:2012ky} is the emergence of the Drude-Sommerfeld
conductivity \eqref{eq:Drude-Sigma} in the low-frequency limit and -- even more
excitingly -- in the mid-frequency range they find scaling  behavior similar to
cuprate superconductors. The use of an explicit lattice may be avoided by using
the simplified form of Homes' law stated in \eqref{eq:Homes-Law-Sum-Rule3} or
\eqref{eq:Homes-Law-Tanner3} which assumes either the validity of the sum rule
or Tanner's law, respectively. This fits in our overall scheme of looking at
universal scaling behavior which should not be affected by an underlying
microscopic lattice.

%%% Local Variables: 
%%% mode: latex
%%% TeX-master: "homes-law"
%%% End: 

\section{Holographic s-Wave Superconductor\label{sec:S-Wave}}

In this section we give a self-contained exposition of the solution to the
equations of motion derived from the Einstein-Maxwell action minimally 
coupled to a charged scalar field. For obtaining a family of holographic
s-wave superconductors, we include the backreaction of the gauge field and the
scalar field on the metric into our analysis. After explaining the normal phase
solutions, i.e.~solutions with a vanishing scalar field, we will employ a
quasi-normal-mode analysis in order to determine the onset of the 
spontaneous condensation of the operator dual to the charged scalar
field. For this purpose, we expand the action up to the quadratic order in the
fluctuations about the background and derive the corresponding equations of
motion. Finally, we determine the R-charge and momentum diffusion related
to the gauge field fluctuations and the metric fluctuations, respectively.
This allows us to calculate the function relevant for testing Homes'
law at finite backreaction.

\subsection{Einstein-Maxwell Action}
The best known (bottom up) holographic model to describe superconductivity is
given by the Einstein-Maxwell action coupled to a charged scalar on the gravity side dual to a field theory with a conserved $U(1)$ current and an
operator describing the condensate \cite{Hartnoll:2008vx,
Hartnoll:2008kx,Herzog:2009xv}. Rescaling the gauge field $A\to eA$ and the
scalar field $\Phi\to e\Phi$ allows us to write the action in the form
\begin{equation}
  S=\frac1{2\kappa^2}\int\dd[\mathit{d+}1]x\sqrt{-g}\left[R-2\Lambda
    -\frac{2\kappa^2}{e^2}\left(\frac14F_{ab}F^{ab}-\abs{\nabla\Phi-\ci A\Phi}^2
      -V(\abs\Phi)\right)\right],
  \label{eq:Einstein-Maxwell-Action}
\end{equation}
where we have a dimensionless coupling parameter
\begin{equation}
  \alpha^2L^2=\frac{\kappa^2}{e^2},
  \label{eq:Definition-Alpha}
\end{equation}
describing the strength of the backreaction onto the geometry exerted by the
gauge field and the scalar field. On the field theory side, this parameter
describes the ratio of charged degrees of freedom to the total degrees of
freedom and thus can be considered as an effective chemical potential or in a
loose sense some kind of ``doping''. The equations of motions for the charged scalar field $\Phi$ derived from \eqref{eq:Einstein-Maxwell-Action} are given by
\begin{equation}
  \begin{aligned}
    \left(\nabla_a+\ci A_a\right)\left(\nabla^a+\ci A^a\right)\cc\Phi-\frac12
    V'(\abs\Phi)\frac{\cc\Phi}{\abs\Phi}&=0, \mnewl
    \left(\nabla_a-\ci A_a\right)\left(\nabla^a-\ci A^a\right)\Phi
    -\frac12V'(\abs\Phi)\frac\Phi{\abs\Phi}&=0.
  \end{aligned}
  \label{eq:Phi-EOM}
\end{equation}
Additionally, we have the Maxwell equations in curved spacetime
\begin{equation}
  \nabla^aF_{ab}=j_b,
  \label{eq:Maxwell}
\end{equation}
which can be simplified to
\begin{equation}
  \frac1{\sqrt{-g}}g_{ab}\partial_c\left(\sqrt{-g}F^{cb}\right)
  =\ci\big[\cc\Phi\left(\nabla_b-\ci A_b\right)
    \Phi-\Phi\left(\nabla_b+\ci A_b\right)\cc\Phi\big].
  \label{eq:Maxwell2}
\end{equation}
Finally, the Einstein equations sourced by the gauge field and by the scalar
field are
\begin{equation}
  R_{ab}-\frac12Rg_{ab}+\Lambda g_{ab}
  =\alpha^2L^2\left(T_{ab}^{\text{em}}+T_{ab}^\Phi\right),
  \label{eq:Einstein}
\end{equation}
with
\begin{align}
  T_{ab}^{\text{em}}&= g^{cd}F_{ac}F_{bd}-\frac14g_{ab}F_{cd}F^{cd},
  \label{eq:Energy-Momentum-EM} \mnewl
  T_{ab}^\Phi&=\left(\nabla_a\cc\Phi+\ci A_a\cc\Phi\right)\left(\nabla_b\Phi
                -\ci A_b\Phi\right)+\left(\nabla_a\Phi-\ci A_a\Phi\right)
             \left(\nabla_b\cc\Phi+\ci A_b\cc\Phi\right) \notag\mnewl
           &\hspace{12pt}-g_{ab}\left(\nabla_c\cc\Phi+\ci A_c\cc\Phi\right)
             \left(\nabla^c\Phi-\ci A^c\Phi\right)-g_{ab}V(\abs\Phi).
  \label{eq:Energy-Momentum-Psi}
\end{align}
The most general planar and rotational symmetric solution to the set of
equations \eqref{eq:Phi-EOM}-\eqref{eq:Einstein} is the AdS-Reissner-Nordström
black hole solution with scalar hair . Note that it is sufficient to assume 
that only quadratic terms are present in the potential
$V(\abs{\Phi})=m^2\abs{\Phi}^2$, since we are only interested in the behavior near the critical point where higher order interactions do not contribute.

\subsection{Normal Phase Solutions of the Background Fields
  \label{ssec:Normal-Phase-S-Wave}}
As we are only interested in the phase diagram and since we are including
the backreaction, it is advisable to simplify our discussion by considering the normal phase background solutions, i.e.~$\Phi\equiv0$, which reduces the equations of motion to
\begin{align}
  R_{ab}-\frac12Rg_{ab}-\frac{d(d-1)}{2L^2}g_{ab}&=\alpha^2L^2T_{ab}^{\text{em}},
  \label{eq:Einstein-Normal} \mnewl
   g^{ab}\nabla_aF_{bc}&=0.
   \label{eq:Maxwell-Normal}
\end{align}
The solutions to these equations are given by
\begin{equation}
  \dd s^2=\frac{L^2}{u^2}\left(-f(u)\dd t^2+\dd{\vvec x}^2
          +\frac{\dd u^2}{f(u)}\right),
  \label{eq:Black-Hole-Metric}
\end{equation}
with
\begin{align}
  f(u)&=1-\bar M\frac{u^d}{u_{\text H}^d}
        +\bar Q^2\frac{u^{2(d-1)}}{u_{\text H}^{2(d-1)}}, &
  \bar M&=1+\bar Q^2, & \bar Q&=\sqrt{\frac{d-2}{d-1}}\mu u_{\text H}\alpha,
  \label{eq:Black-Hole-Definitions}
\end{align}
where $\bar M$ and $\bar Q$ denotes the dimensionless mass and the
dimensionless charge of the black hole, respectively. The gauge field solution
including the constraints on the black hole horizon may be written as
\begin{equation}
  A=\mu\left(1-\frac{u^{d-2}}{u_{\text H}^{d-2}}\right)\dd t,
  \label{eq:Gauge-Field-Normal}
\end{equation}
where $\mu$ is the chemical potential, describing the difference in the 
electric potential between the horizon of the black hole at $u=u_{\text H}$ and
the boundary of the AdS space (at $u=0$) and thus the energy of adding another
charged particle to the black hole. Furthermore we may define a temperature
related to the Bekenstein entropy by
\begin{equation}
  T=\frac1{2\pi}\left[\frac1{\sqrt{g_{uu}}}\dev u\sqrt{-g_{tt}}
     \right]_{u=u_{\text H}}=\frac{d-(d-2)\bar Q^2}{4\pi u_{\text H}}.
   \label{eq:Black-Hole-Temperature}
\end{equation}
Since we are dealing with a scale invariant theory in the UV, the only
dimensionless physical parameter is the ratio $\nicefrac T\mu$ which will be
controlled (numerically) by the dimensionless chemical potential
$\bar\mu=\mu u_{\text H}$. Using \eqref{eq:Black-Hole-Definitions} we may write
this ratio in terms of the charge of the black hole,
\begin{equation}
  \frac T\mu=\frac{d-\frac{(d-2)^2}{d-1}\bar\mu^2\alpha^2}{4\pi\bar\mu}
  =\frac{d-(d-2)\bar Q^2}{\frac{4\pi}\alpha\sqrt{\frac{d-1}{d-2}}\bar Q},
  \label{eq:ToverMu}
\end{equation}
and conversely
\begin{equation}
  \bar Q_\pm=\frac{-2\pi\sqrt{d-1}\left(\frac T\mu\right)
    \pm\sqrt{4\pi^2(d-1)\left(\frac T\mu\right)^2+d(d-2)^2\alpha^2}}
  {\alpha(d-2)^{\frac32}}.
  \label{eq:Q-In-Terms-Of-ToverMu}
\end{equation}
In order to be consistent with the probe limit $\alpha\to0$, we pick the 
positive solution of the quadratic equation for the black hole charge $\bar Q$,
such that $\lim_{\alpha\to0}\bar Q_+=0$.

\subsection{Quasi-Normal-Mode Analysis and Phase Diagram}
\begin{figure}[t]
  \centerline{
    \includegraphics[width=.48\textwidth]{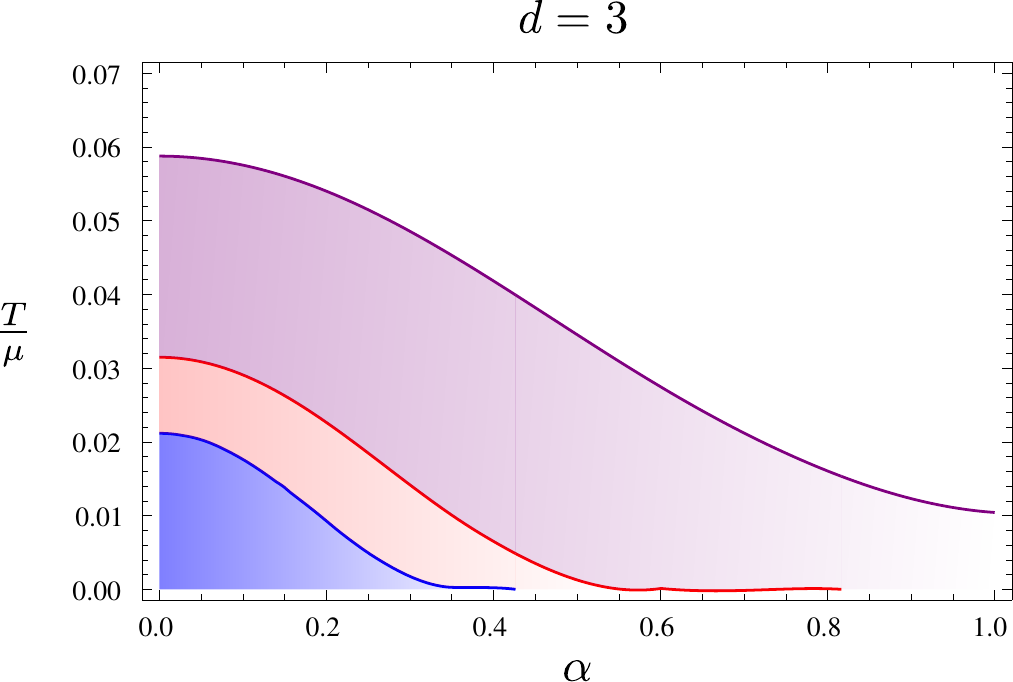}
    \hspace{.04\textwidth}
    \includegraphics[width=.48\textwidth]{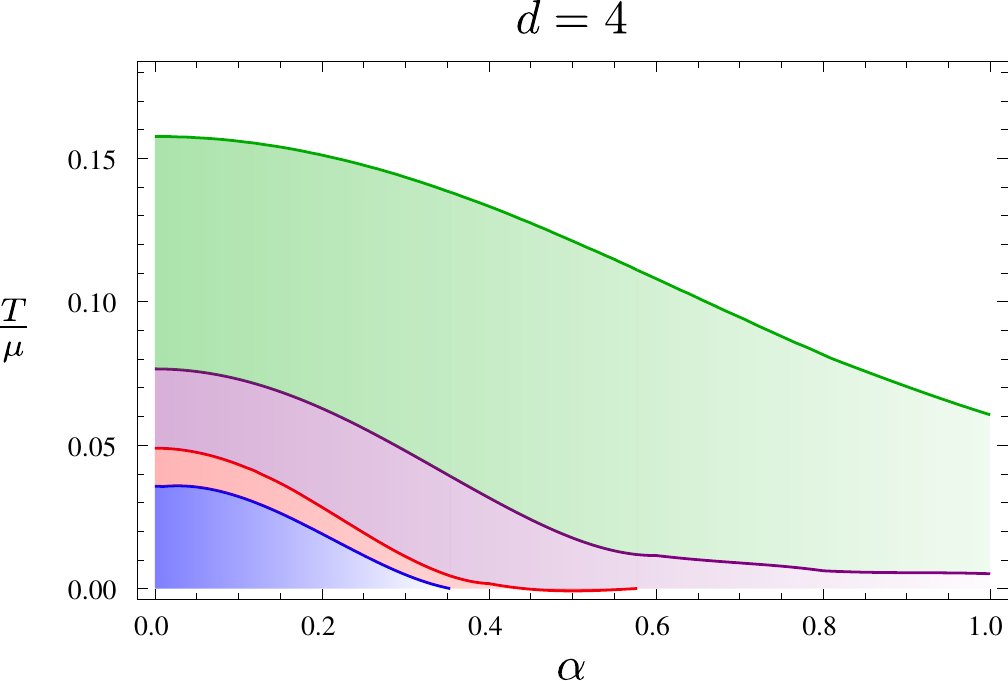}
   }
  \caption{Phase diagram of the holographic s-wave superconductor for $d=4$ 
    and $d=3$ as a function of the backreaction parameter $\alpha$, 
    depending on the scalar field mass. Colored regions (for $d=3$
    are encoded as $\color{blue}m^2L^2=4$, $\color{red}m^2L^2=0$, 
    $\color{mpurple}m^2L^2=-2$ and for $d=4$ we have $\color{blue}m^2L^2=5$, 
    $\color{red}m^2L^2=0$, $\color{mpurple}m^2L^2=-3$,  
    $\color{mdarkgreen}m^2L^2=-4$) show phases where the scalar field
    condenses yielding a superfluid phase.\label{fig:Phase-Diag-S-Wave}}
\end{figure}

We use the holographic dissipation-fluctuation theorem \cite{Son:2002sd} in
order to determine the complex-valued Green's function of scalar fluctuations
about the fixed scalar background in the normal phase. Therefore we derive the
quadratic action in the fluctuations
\begin{equation}
  \begin{aligned}
    \phi(t,\vvec x,u)&=\Phi(u)+\delta\phi(t,\vvec x,u), \mnewl
    A_a(t,\vvec x,u)&=A_a^{\text B}(u)+a_a(t,\vvec x,u), \mnewl
    g_{ab}(t,\vvec x,u)&=G_{ab}(u)+h_{ab}(t,\vvec x,u).
  \end{aligned}
  \label{eq:Ansatz-Fluctuations}
\end{equation}
The detailed calculations and solutions are shown in Appendix
\ref{sec:EOM-Fluctuations}. For convenience we state the equation of motion for
the scalar field fluctuations \eqref{eq:Scalar-Fluctuations-Appendix},
\begin{equation}
  \delta\phi''(u)+\left(\frac{f'(u)}{f(u)}-\frac{d-1}u\right)\delta\phi'(u)
  +\left[\frac{(\omega+A_t)^2}{f(u)^2}-\frac{\vvec k^2}{f(u)}
    -\frac{L^2m^2}{u^2f(u)}\right]\delta\phi(u)=0,
  \label{eq:Scalar-Fluctuations}
\end{equation}
which we will use in order to determine the phase diagram. We are interested in
the spontaneous symmetry breaking induced by the condensation of the operator
dual to the scalar field $\Phi$. For a second order phase transition we can use
the dissipation-fluctuation theorem to look for instabilities in the
fluctuations. Thus, for a given fixed value of the backreaction $\alpha$, we 
vary the dimensionless parameter $\nicefrac T\mu$ and look for a quasi-normal
mode crossing the origin of the complex frequency plane. Numerically, we solve
\eqref{eq:Scalar-Fluctuations} with infalling boundary conditions at the black
hole horizon and calculate the corresponding retarded Green's function for
$\vvec k=\vvec0$ and $\omega=0$. We expect to find a gapless mode which is
related to a pole in the retarded Green's function, such that we may determine the set of parameters $(\alpha,\nicefrac T\mu)$ yielding critical curves as shown in \figref{fig:Phase-Diag-S-Wave}.
Additionally, we can also vary the mass of the scalar field related to the
scaling dimension of the dual operator via
\begin{alignat}{2}
  \Delta_\pm&=\frac12\left(d\pm\sqrt{d^2+4L^2m^2}\right), &\qquad\qquad 
  m^2L^2&=\Delta_\pm\left(\Delta_\pm-d\right).
  \label{eq:Scaling-Relation-Scalar-Field}
\end{alignat}
The critical value of the backreaction for $T=0$ can be determined analytically
by looking at the IR behavior near the black hole horizon. The corresponding
charge of the black hole is given by
\begin{equation}
  \bar Q^2=\frac d{d-2}.
  \label{eq:Critical-Qbar}
\end{equation}
Inserting this value into the metric \eqref{eq:Black-Hole-Metric}, we find a
double zero giving rise to an $\text{AdS}_2\times\mathds R^{d-1}$ metric and
thus an IR fixed point
\begin{equation}
  \dd s^2_{\text{AdS}_2}=L^2\left(\varrho^2\dd{\tilde t}
                      +\frac{\dd\varrho^2}{\varrho^2}\right),
  \label{eq:AdS2-Rd-1-Metric}
\end{equation}
with $\varrho=(u-u_{\text H})$ and rescaled $t$ and $\vvec x$
coordinates, accordingly. The $\text{AdS}_2$ radius is related to the
$\text{AdS}_{d+1}$ radius by
\begin{equation}
  L_{\text{AdS}_2}^2=\frac{L^2}{d(d-1)}.
  \label{eq:Relation-AdS2-Curvature}
\end{equation}
The near horizon expansion of the background gauge field reads
\begin{equation}
  A_t^2\approx\frac{d(d-1)}{u_{\text H}^4\alpha^2}(u-u_{\text H})^2
  =\frac{d(d-1)}{u_{\text H}^4\alpha^2}\varrho^2.
  \label{eq:Near-Horizon-Expansion-Phi2}
 \end{equation}
Finally, \eqref{eq:Scalar-Fluctuations} can be written as
\begin{equation}
  \delta\psi''(\varrho)+\frac2\varrho\delta\psi'(\varrho)
  -\frac{L^2_{\text{AdS}_2}m^2}{\varrho^2}\delta\psi(\varrho)=0,
  \label{eq:Scalar-Field-AdS2}
\end{equation}
and the effective mass term can be compared to the Breitenlohner-Freedman
stability bound for $\text{AdS}_2$,
\begin{equation}
  L^2_{\text{AdS}_2}m^2_{\text{eff}}=\frac1{d(d-1)}
  \left(L^2m^2-\frac1{\alpha^2}\right)\leq L^2_{\text{AdS}_2}m^2_{\text{BF}}=-\frac14.
  \label{eq:Backreaction-Stability-Bound}
\end{equation}
This leads to a condition on $\alpha$ for scalar field condensation to
occur,
\begin{equation}
  \alpha^2\leq\frac1{\frac{d(d-1)}4+L^2m^2}.
  \label{eq:Critical-Alpha-T0}
\end{equation}
We see if the mass is already below the $\text{AdS}_2$ Breitenlohner-Freedman
bound, there is no critical value for $\alpha$ and hence no quantum critical
point or phase transition at zero temperature between the condensed phase and
the normal phase. The different masses used in the numerical calculation and 
the corresponding values for the scaling and the critical backreaction strength
are listed in \tabref{tab:Critical-Alpha}.
\begin{table}[t]
  \centering
  \begin{tabular}{@{}T{2cm}S{1cm}S{1cm}S{2cm}S{1cm}S{1cm}S{1cm}C@{}}
    \toprule
    & \multicolumn{3}{S{22mm}}{d=3} & \multicolumn{4}{C}{d=4} \\ \midrule
    m^2L^2 & 4 & 0 & -2 & 5 & 0 & -3 & -4 \\[2mm]
    \Delta_- & -1 & 0 & 1 & -1 & 0 & 1 & 2 \\
    \Delta_+ & 4 & 3 & 2 & 5& 4 & 3 & 2 \\[2mm]
    \alpha^2_c & \dfrac2{11} & \dfrac23 & -2 
    & \dfrac18 & \dfrac13 & \infty & -1 \\[2mm]
    \bottomrule
  \end{tabular}
  \caption{List of the critical value for the strength of the backreaction
    $\alpha$ for different masses in three and four dimensions. Note that the
    instability condition is satisfied for $\alpha<\alpha_c$. In particular for
    the negative values we do not have a critical value of
    $\alpha\in\mathds R$, so in this case for $T=0$ we always find a 
    condensed/superfluid phase.
    \label{tab:Critical-Alpha}}
\end{table}
Our results as displayed in \figref{fig:Phase-Diag-S-Wave}
show that the critical temperature decreases with increasing backreaction
strength $\alpha$. Moreover, if the scalar mass is larger than a critical 
value, the critical temperature goes to zero for a finite value of
$\alpha$. This is the case most reminiscent of a real high $T_c$ superconductor,
when the dome in \figref{fig:Phase-Diag-S-Wave} has similarities with the
right hand side of the dome in the phase diagram of a high $T_c$ superconductor.
\\[\baselineskip]
The physical interpretation of $\alpha$ is that it corresponds to the ratio of
the number of $SU(2)$ charged degrees of freedom over all degrees of freedom
\cite{Ammon:2009xh}. The phase diagrams above indicate that an increase of
$\alpha$ reduces the numbers of degrees of freedom which can participate in 
pair formation and condensation, such that $T_c$ is lowered. A similar mechanism
also seems to be at work when adding a double trace deformation to the
holographic superconductor \cite{Faulkner:2010gj}, and has been discussed 
within condensed matter physics using a BCS approach in \cite{Zaanen:2009prb}. For holographic superconductors, this mechanism is most clearly visible for the
top-down holographic superconductors involving D7 brane probes \cite{Ammon:2008fc,Ammon:2009fe} in which the dual field theory Lagrangian and
thus the field content of the condensing operator are known. In these models,
there is an $U(2)$ symmetry which factorizes into an $SU(2)_I\times U(1)_B$,
i.e.~into an isospin and a baryonic symmetry. A chemical potential is switched
on for the $SU(2)_I$ isospin symmetry and the condensate is of $\rho$-meson
form,
\begin{equation}
  J_x{}^1=\bar\psi\sigma^1\gamma_x\psi+\bar\phi\sigma^1\partial_x\phi 
         =\bar\psi_\uparrow\gamma_x\psi_\downarrow
          +\bar\psi_\downarrow\gamma_x\psi_\uparrow+\text{bosons},
  \label{eq:Rho-Meson-Condensate}
\end{equation}
where $\psi=(\psi_\uparrow, \psi_\downarrow)$ and $\phi=(\phi_\uparrow,\phi_\downarrow)$ are the quark and squark doublets,
respectively, which involve up and down flavors, with $\sigma^i$ the Pauli
matrices in isospin space and $\gamma_\mu$ the Dirac matrices. As an additional
control parameter, a chemical potential $\mu_B$ for the baryonic $U(1)_B$
symmetry may be turned on. This leads to a decrease of $T_c$
\cite{Erdmenger:2011hp} (see also \cite{Bigazzi:2011ak}) which may be 
understood as follows: Under the $U(1)_B$ symmetry, $\psi$ and ${\bar \psi}$
have opposite charge. The same applies to $\phi$ and $\bar \phi$. Turning on
$\mu_B$ leads to an excess of $\psi$ over $\bar \psi$ degrees of freedom,  which
implies that less degrees of freedom are available to form the pairs
\eqref{eq:Rho-Meson-Condensate}. The same applies to $\phi$ and $\bar\phi$
degrees of freedom. Thus in this case, charge carriers in the normal state are
also present in the superconducting phase, leading to the formation of a
pseudo-gap.

\subsection{Diffusion Constants}
As explained in Section \ref{sec:Homes-Law}, an analysis of Homes' law requires
the definition of a relevant time scale. Possible candidates for time scales 
are associated with diffusive processes. We may  determine two different types
of diffusion, momentum diffusion $D_{\text M}$ and R-charge diffusion
$D_{\text R}$, respectively. The former can be related to the shear viscosity by
studying hydrodynamic modes,
\begin{equation}
  D_{\text M}=\frac\eta{\varepsilon+P},
  \label{eq:Momentum-Diffusion}
\end{equation}
such that it can be solely described by thermodynamic quantities using the 
famous result \eqref{eq:Eta-Over-S}. The latter, however, cannot be calculated
solely by thermodynamic quantities in general, such that an independent
calculation for each background is necessary. Usually the charge diffusion
constant can be read off from the dispersion relation $\omega=-\ci D\vvec k^2$
of the conserved current, which can be derived from Fick's law and the continuity equation
\begin{alignat}{5}
  \vvec j&=-D\nabla\rho, &\qquad &\text{and} &\qquad
  \dot\rho+\nabla\scp\vvec j&=0, &\qquad\quad &\longrightarrow &\qquad\quad
  \dot\rho-D\nabla^2\rho&=0.
  \label{eq:Diffusion-Equation}
\end{alignat}
Fortunately, this work has already been done by \cite{Son:2006em} without backreaction.

\subsubsection*{Momentum Diffusion with Backreaction}
Since we are working with a fixed chemical potential, we take the grand 
canonical ensemble to describe the momentum diffusion in the thermodynamic
limit. Using the gauge/gravity duality the grand canonical potential is given by
\begin{equation}
  \Omega=\frac1\beta I=TI,
  \label{eq:Grand-Canonical-Potential}
\end{equation}
where $I$ denotes the regularized Euclidean on-shell action given by
\begin{equation}
  I=-\frac{L^{d-1}}{2\kappa^2}\frac{V_{d-1}}T\frac1{u_{\text H}^d}
     \left(1+\bar Q^2\right).
\end{equation}
Using the Gauge/Gravity dictionary, we derive the charge density, the energy
density and the entropy density from \eqref{eq:Grand-Canonical-Potential}, with the help of the relations
\begin{equation}
  \begin{aligned}
    s&=\frac{4\pi}{2\kappa^2}\frac{L^{d-1}}{u_{\text H}^{d-1}}, \mnewl
    \varepsilon&=\frac{d-1}{2\kappa^2L^{d-1}}\frac{1+\bar Q^2}{u_{\text H}^d},
    \mnewl
    n&=\sqrt{(d-1)(d-2)}\frac{L^{d-1}}{\kappa^2}
       \frac{\alpha\bar Q}{u_{\text H}^{d-1}}.
     \end{aligned}
  \label{eq:Field-Theory-SEN-Normal-Backreaction}
\end{equation}
The black hole horizon $u_{\text H}$ should be understood as a function of $\mu$
or $T$, respectively, defined by solving either \eqref{eq:Black-Hole-Definitions} or \eqref{eq:Black-Hole-Temperature} for
$u_{\text H}$, whereas $\bar Q$ is a function of $\nicefrac T\mu$ defined by
the inversion of \eqref{eq:Q-In-Terms-Of-ToverMu}. Starting from the definition
of the momentum diffusion and using the relations \eqref{eq:Eta-Over-S} and the
first law of thermodynamics $\varepsilon=Ts-P+\mu n$, we find
\begin{equation}
  D_{\text M}=\frac\eta{\varepsilon+P}=\frac{\frac s{4\pi}}{Ts+\mu n}
           =\frac1{4\pi T}\frac1{1+\frac\mu T\frac ns}.
  \label{eq:Momentum-Diffusion2}
\end{equation}
Inserting the relations \eqref{eq:ToverMu} for $\nicefrac\mu T$ and \eqref{eq:Field-Theory-SEN-Normal-Backreaction} for
$\nicefrac ns$ we finally arrive at
\begin{equation}
  D_{\text M}=\frac1{4\pi T}\inv{\left(1
            +\frac{2(d-1)\bar Q^2}{d-(d-2)\bar Q^2}\right)}.
  \label{eq:Momentum-Diffusion3}
\end{equation}

\subsubsection*{R-Charge Diffusion with Backreaction}
As explained in the previous subsection it is not straightforward to generalize
the R-charge diffusion with backreaction to arbitrary dimensions. To our
knowledge the functional dependence on the backreaction of $D_{\text R}$ in
$d=3$ dimensions is not known yet, let alone for arbitrary dimensions
$d$. Surely, this would be an interesting result which is beyond the scope of this work. In $d=4$ the R-charge diffusion with backreaction is determined in
\cite{Banerjee:2008th}. Converting the expression for $D_{\text R}$ into the
form given by our conventions we find
\begin{equation}
  D_{\text R}=\frac1{4\pi T}\frac{(2-\bar Q^2)(2+\bar Q^2)}{2(1+\bar
    Q^2)}\, ,
  \label{eq:R-Charge-Diffusion-Backreaction-4d}
\end{equation}
where the dimensionless black hole charge is defined in
\eqref{eq:Black-Hole-Definitions}. Furthermore, we may look at the ratio of the
two different diffusion constants,
\begin{equation}
  \frac{D_{\text R}}{D_{\text M}}=2+\bar Q^2.
  \label{eq:Ratio-Diffusion-Constants}
\end{equation}

\subsection{Discussion\label{ssec:Discussion}}

\subsubsection*{Diffusion without backreaction for arbitrary $d$}
\begin{figure}[t]
  \centerline{
    \includegraphics[width=.48\textwidth]{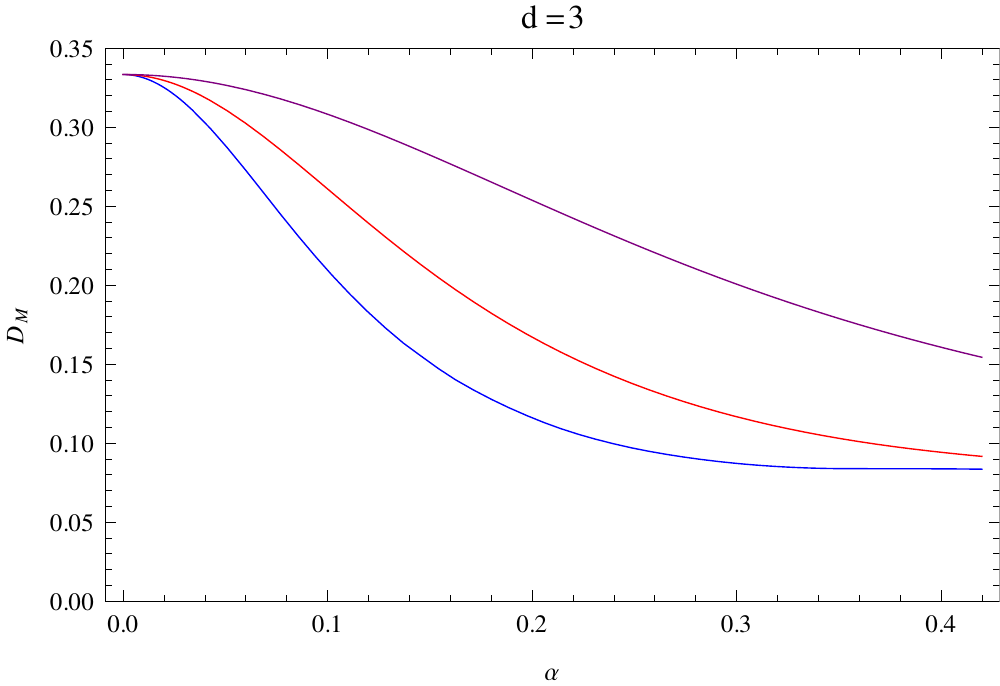}
    \hspace{.04\textwidth}
    \includegraphics[width=.48\textwidth]{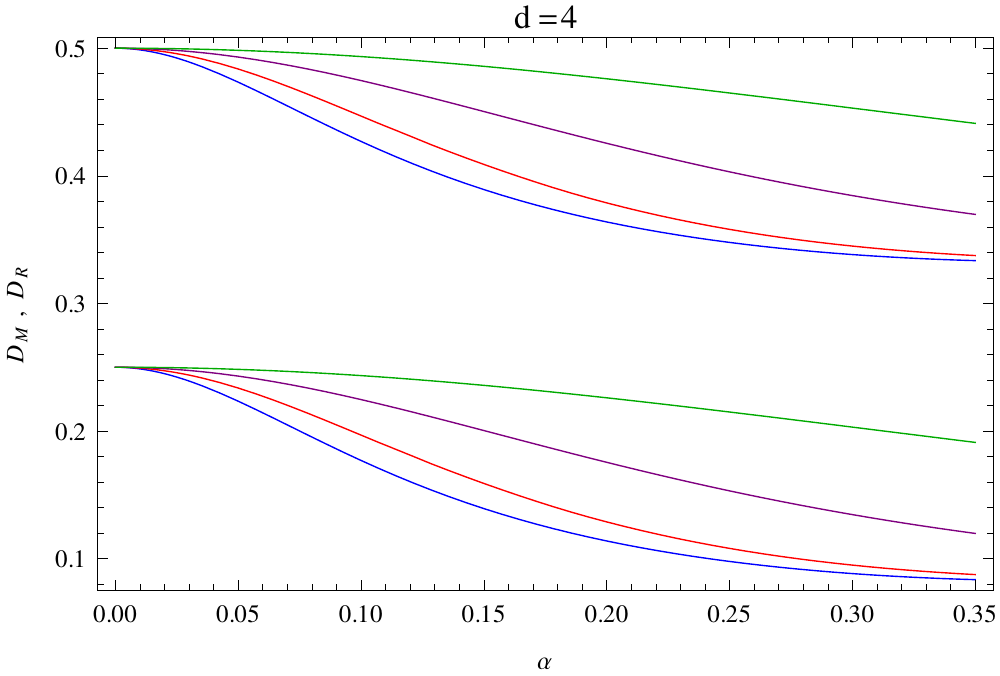}
  }
  \caption{Plots of momentum diffusion constant $D_{\text M}$ for $d=3$ (left
    panel) and R-charge diffusion $D_{\text R}$ and momentum diffusion
    $D_{\text M}$ for $d=4$ (right panel) versus the backreaction strength
    $\alpha$. The color coding of the different scalar field masses follows 
    \figref{fig:Phase-Diag-S-Wave}. The probe limit properties for $\alpha=0$ 
    are reproduced, i.e.~for $d=3$ we find $D_{\text M}=\nicefrac13$ and in the 
    case of $d=4$ $D_{\text M}=\nicefrac14$ and $D_{\text R}=\nicefrac12$, 
    independently of the scalar field mass.\label{fig:Diffusion-Constants}}
\end{figure}
First of all we check our results by taking the limit of vanishing
backreaction, i.e.~setting $\alpha$ and $\bar Q$ to zero in
\eqref{eq:Momentum-Diffusion3} and
\eqref{eq:R-Charge-Diffusion-Backreaction-4d}. We see that the dimensionless
temperature $\bar T$ \eqref{eq:Black-Hole-Temperature} has the fixed value of
$\nicefrac d{4\pi}$ and thus
\begin{equation}
  \bar D_{\text M}=\frac1{4\pi\bar T}=\frac1d,
  \label{eq:Momentum-Diffusion-No-Backreaction}
\end{equation}
is equal for all masses of the scalar field. Similarly, the R-charge diffusion
is given by \cite{Kovtun:2008kx}
\begin{equation}
  \bar D_{\text R}=\frac1{4\pi\bar T}\frac d{d-2}=\frac1{d-2},
  \label{eq:R-Charge-Diffusion}
\end{equation}
and the ratio is given by
\begin{equation}
  \frac{D_{\text R}}{D_{\text M}}=\frac d{d-2}.
  \label{eq:Ratio-Diffusion-Constants-No-Backreaction}
\end{equation}
Comparing these results to our numerical solutions shown in
\figref{fig:Diffusion-Constants}, we see that for $\alpha=0$, we
obtain for $d=3$ dimensions a dimensionless value of $\nicefrac13$ for 
$D_{\text M}$. This is consistent with the values obtained form the analytic
calculation without backreaction \eqref{eq:Momentum-Diffusion-No-Backreaction}. In particular, we
see  that the momentum diffusion does not depend on the mass of the
background scalar field since all diffusion constants for different masses
(indicated in the figure by different colors) converge to the same value.
Similarly for $d=4$, we check our numerical values for the momentum and the
R-charge diffusion by comparing to the results without backreaction. As is
displayed in \figref{fig:Diffusion-Constants}, we find that
$D_{\text M}=\nicefrac14$ and $\bar D_{\text R}=\nicefrac12$ as stated in
\eqref{eq:Momentum-Diffusion-No-Backreaction} and \eqref{eq:R-Charge-Diffusion}, as well as the correct value of the ratio
$\nicefrac{D_{\text R}}{D_{\text M}}=2$. Again our numerical results are
consistent with the analytic solutions without backreaction and we see the same
convergence effect for the different masses of the scalar field for each
diffusion constant displaying the independence on the scalar fields' mass. 

\subsubsection*{Momentum Diffusion for $d=3$}
\begin{figure}[t]
  \centerline{
    \includegraphics[width=.48\textwidth]{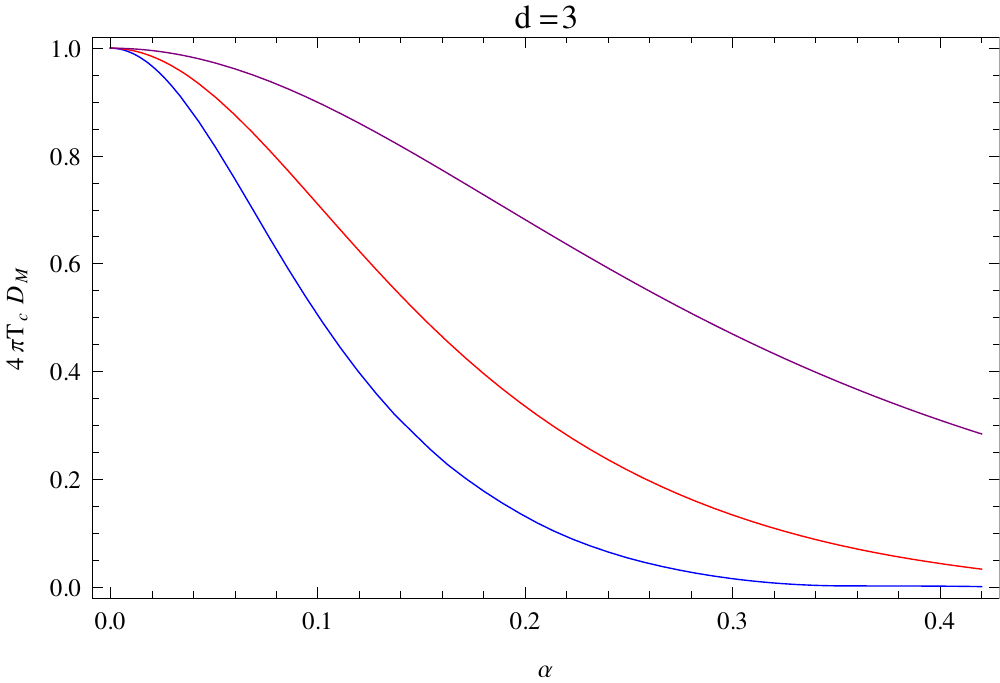}
    \hspace{.04\textwidth}
    \includegraphics[width=.48\textwidth]{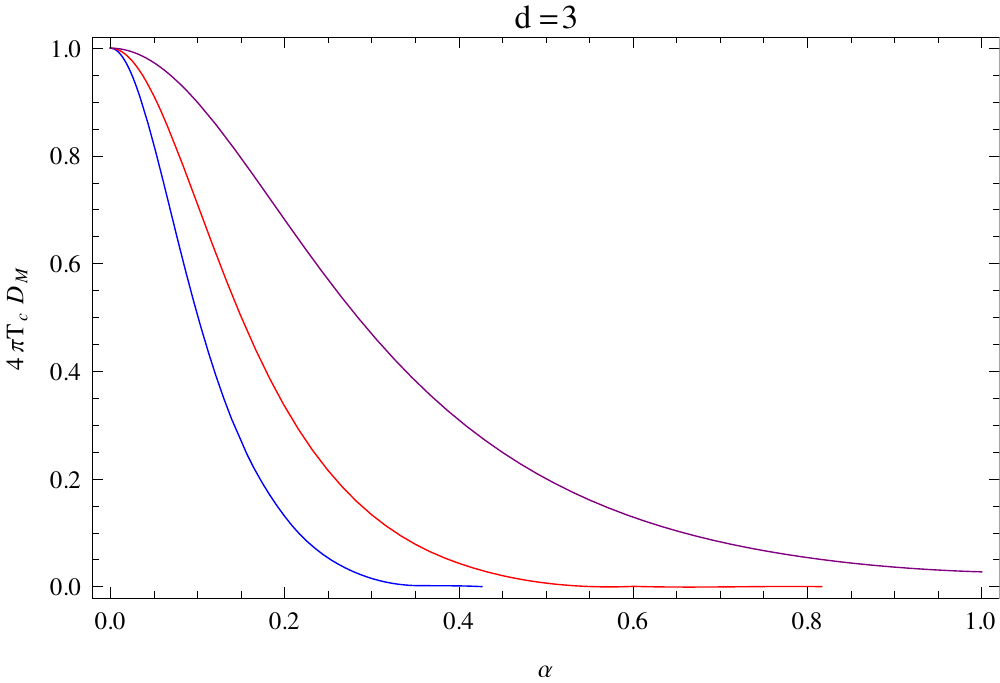}
  }
  \caption{The constant $C_{\text M}$ related to momentum diffusion $D_{\text M}$ 
    for $d=3$ and different masses of the scalar field (color coded as
    $\color{blue}m^2L^2=4$, $\color{red}m^2L^2=0$ and 
    $\color{mpurple}m^2L^2=-2$) plotted depending on the strength of
    the backreaction $\alpha$, defined in \eqref{eq:Definition-Alpha}. The 
    left panel shows the numerical values found by a minimization algorithm 
    up to the $\alpha$ where poles of higher excitations in the Green's 
    function overlays with the lowest modes, thus leading to a breakdown of the
    algorithm. Only in the case of $\color{blue}m^2L^2=-2$, we have tested
    our algorithm up to $\alpha=1$. In the right panel we included the
    analytically determined critical points $\alpha_c$ for 
    $\color{blue}m^2L^2=4$ and $\color{red}m^2L^2=0$ given in
    \tabref{tab:Critical-Alpha} which corresponds to the zero temperature case
    with $\bar Q^2=\nicefrac d{(d-2)}$ and thus $C_{\text M}=0$. Due to the
    reduced number of data points we use an interpolation function in the 
    region $\alpha\in[0.35,\sqrt{\nicefrac23}\approx0.82]$. For this
    reason the curves for $\color{blue}m^2L^2=4$ and $\color{red}m^2L^2=0$
    should not be trusted to be very accurate. Since for 
    $\color{mpurple}m^2L^2=-2$ there is no critical value of $\alpha$, so 
    $C_{\text M}$ approaches zero asymptotically. 
    \label{fig:C-Constants-3D}}
\end{figure}

Inserting the critical values for $\nicefrac T\mu$ at a fixed
$\alpha$, we may check how much the momentum diffusion is varying with respect 
to $\alpha$. As shown in \figref{fig:C-Constants-3D}, there is a
mild variation of the momentum diffusion in the sense that we are approaching
the values for $\alpha=0$ in a linear way.
In order to compare our numerical results to the constant governing Homes' law
rewritten in the form \eqref{eq:Homes-Law-Sum-Rule3} or \eqref{eq:Homes-Law-Tanner3}, we introduce the function
\begin{equation}
  C_{\text M}=4\pi T_cD_{\text M}(T_c)
           =\inv{\left(1+\frac{4\bar Q^2}{3-\bar Q^2}\right)},
  \label{eq:C-Constant-3D}
\end{equation}
plotted in \figref{fig:C-Constants-3D}. Here we assume that the
diffusive process is proportional to a time scale 
$\tau_c\propto D_{\text M}(T_c)$, such that the proportionality does not depend
on $\bar Q$ or $\alpha$. 

\subsubsection*{R-Charge Diffusion and Momentum Diffusion for $d=4$ }
We now repeat the discussion of the previous subsection for the four-dimensional case. In addition to the momentum diffusion, we may also
calculate the R-charge diffusion which will give us an additional time scale to choose.
\begin{figure}[t]
  \centerline{
    \includegraphics[width=.48\textwidth]{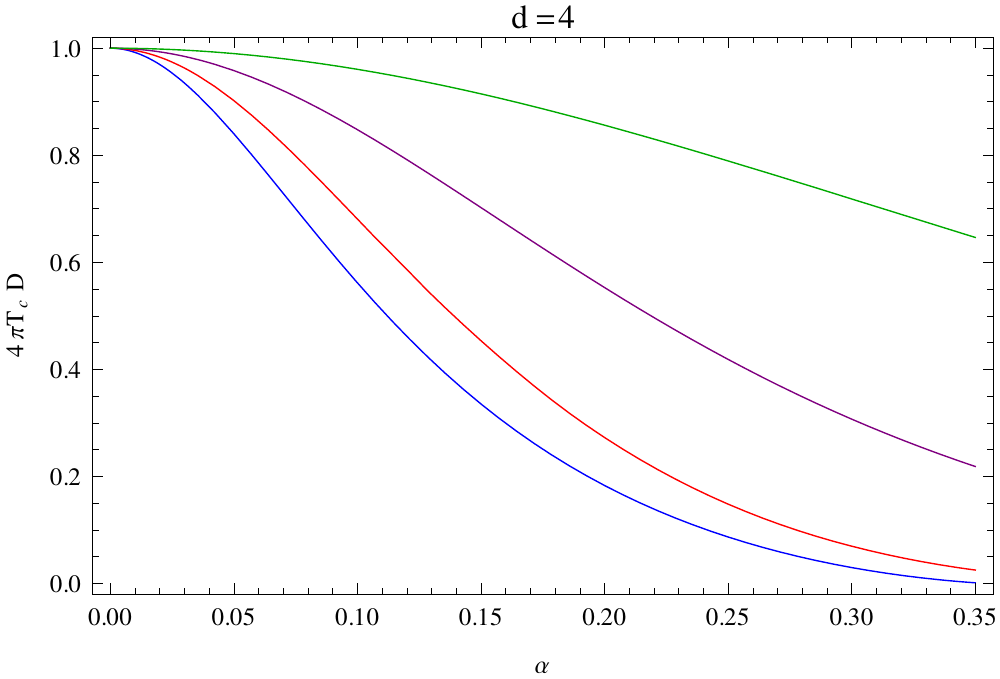}
    \hspace{.04\textwidth}
    \includegraphics[width=.48\textwidth]{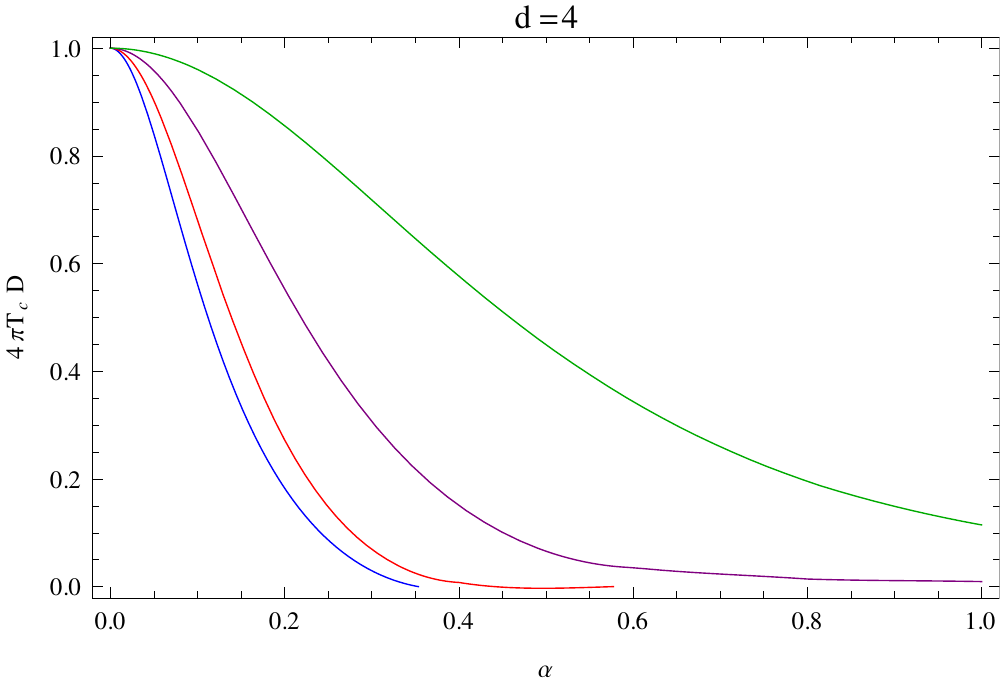}
  }
  \caption{For $d=4$ dimensions: Dimensionless momentum 
    diffusion given by $C_{\text M}$ depending on the backreaction $\alpha$ (for
    different masses of the scalar field color coded as 
    $\color{blue}m^2L^2=5$, $\color{red}m^2L^2=0$, $\color{mpurple}m^2L^2=-3$ 
    and $\color{mdarkgreen}m^2L^2=-4$). As already explained in 
    \figref{fig:C-Constants-3D}, the endpoint in the curves for scalar field
    mass $\color{blue}m^2L^2=5$ and $\color{red}m^2L^2=0$ as shown in the right
    panel is set by the critical values of $\alpha$ listed in
    \tabref{tab:Critical-Alpha}. Due to the interpolation these curves should 
    be taken with a grain of salt in the vicinity of the critical point 
    $\alpha_c$.\label{fig:CM-Constants-4D}}
\end{figure}
The constant related to $4\pi\tau_cT_c$ for the R-charge diffusion reads
\begin{align}
  C_{\text R}&=\frac{(2-\bar Q^2)(2+\bar Q^2)}{2(1+\bar Q^2)}.
  \label{eq:CR-Constant-4D}
  \intertext{Note that this function is running from $2$ to $0$ for
    $\alpha\in[0,\alpha_c]$ since the R-charge diffusion without backreaction is
    given by $\nicefrac1{(2\pi T)}$ see \eqref{eq:R-Charge-Diffusion}. For
    $d=4$, the constant for the momentum diffusion reads}
  C_{\text M}&=\inv{\left(1+\frac{3\bar Q^2}{2-\bar Q^2}\right)}.
  \label{eq:CM-Constant-4D}
\end{align}
Note also that the ratio $\nicefrac{C_{\text R}}{C_{\text M}}$ is given by $2$ in
compliance with the ratio for the diffusion constants
\eqref{eq:Ratio-Diffusion-Constants-No-Backreaction}. The plots of $C_{\text M}$
and $C_{\text R}$ are shown in \figref{fig:CM-Constants-4D} and
\figref{fig:CR-Constants-4D}, respectively.
\begin{figure}[t]
  \centerline{
    \includegraphics[width=.48\textwidth]{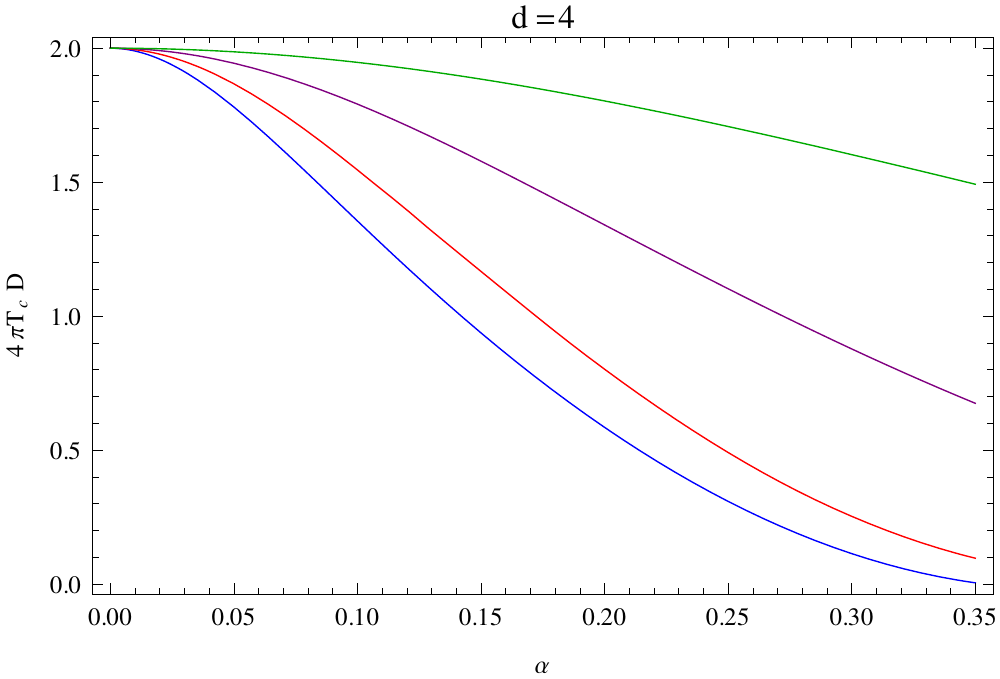}
    \hspace{.04\textwidth}
    \includegraphics[width=.48\textwidth]{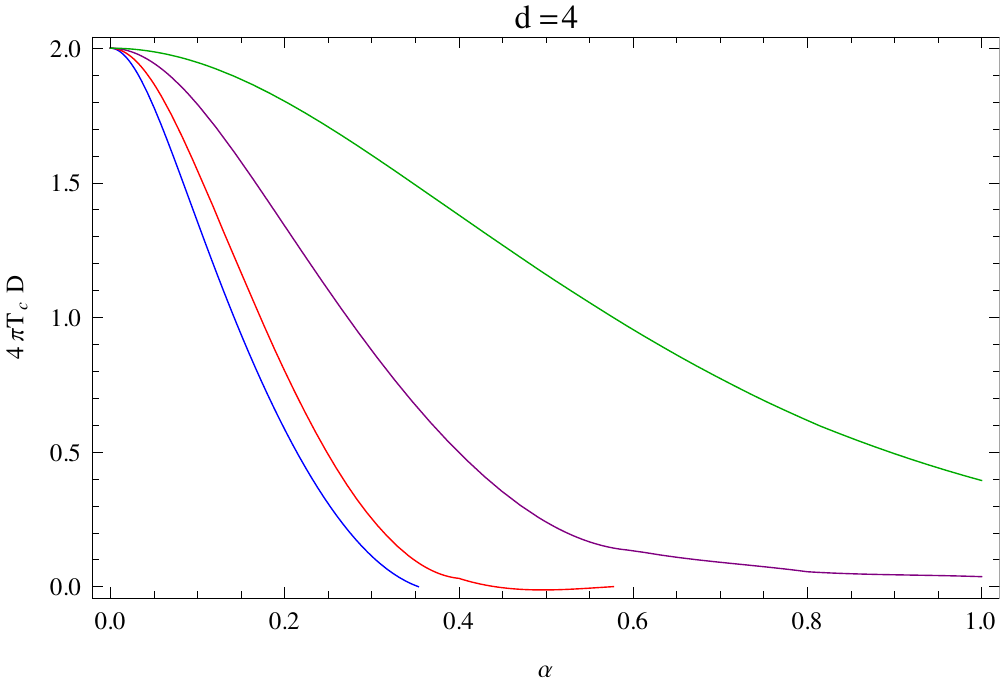}
  }
  \caption{Dimensionless R-charge diffusion given by
    $C_{\text R} = 4 \pi T_c D_{\text R}$. Note that we have used the same
    prefactor in the definition \eqref{eq:CR-Constant-4D} as for the 
    dimensionless momentum $C_{\text M}$, which leads to the probe limit value
    of two. The color coding of the scalar field masses is identical to 
    \figref{fig:CM-Constants-4D}.\label{fig:CR-Constants-4D}}
\end{figure}
We see that the numerically solutions are virtually the same as in the three
dimensional case. In both dimensions the numerical solutions display the same
properties:
\begin{itemize}

\item Instead of the behavior $D(T_c) T_c = const.$ predicted by
  Homes' law in combination with the sum rules as discussed in
  Section \ref{ssec:Holographic-Homes-Law}, we observe a decrease of this
  quantity with increasing backreaction strength $\alpha$. We discuss possible
  explanations for this behavior below in section 5.
\item A test on our calculations is provided by the fact that 
numerically, we reproduce the results known from probe limit calculations
  where by definition $C_{\text M}$ and $C_{\text R}$ are one or two, respectively.
\item The region where our minimization algorithm is working safely is
  approximately given by $\alpha\approx0.35$. For higher values we use an
  interpolation function for the curves associated to the non-negative 
  scalar field masses including the zero temperature values
  obtained analytically.
\item At the endpoints where we encounter a quantum critical point with
  vanishing critical temperature, $C_{\text M}$ and $C_{\text R}$ are
  zero since these points correspond to the critical $\alpha$ stated in
  \tabref{tab:Critical-Alpha}. For the negative masses there exists no critical
  $\alpha$ and the curves are reaching zero only asymptotically.
\end{itemize}
%

%%% Local Variables: 
%%% mode: latex
%%% TeX-master: "homes-law"
%%% End: 

\section{Holographic p-Wave Superconductor\label{sec:P-Wave}}

In this section we discuss the Einstein-Yang-Mills theory including 
backreaction in asymptotically $AdS_5$ which give rise to a holographic p-wave
superfluid at low temperatures \cite{Ammon:2009xh}. In addition to the
Abelian symmetry the rotational symmetry $SO(3)$ is spontaneously
broken down to $SO(2)$. As far as Homes' law is concerned, we observe
very similar dependence of $D(T_c) T_c$ on the backreaction as in the
s-wave case.

\subsection{Einstein-Yang-Mills Action
  \label{ssec:Einstein-Yang-Mills-Action}}
In addition to gravity, we consider here gauge fields transforming under some
gauge group. This system is described by the Einstein-Yang-Mills action. In the
following we  specialize to $(4+1)$-dimensional asymptotically AdS space and to the gauge group $SU(2)$ with field strength tensor
\begin{equation}
  F^a_{\mu\nu}=\partial_\mu A^a_\nu-\partial_\nu A^a_\mu 
             +\epsilon^{abc}A^b_\mu A^c_\nu,
  \label{eq:FSU(2)}
\end{equation}
where $\epsilon^{abc}$ is the total antisymmetric tensor and $\epsilon^{123}=+1$. In analogy with QCD, we call the $SU(2)$ isospin symmetry. 
\\[\baselineskip]
The Einstein-Yang-Mills action reads
\begin{equation}
  S=\int\dd[5]x \sqrt{-g}\left[\frac{1}{2\kappa_5^2}(R-\Lambda)
    -\frac{1}{4\hat{g}_\text{YM}^2}F^a_{\mu\nu}F^{a\mu\nu}\right] \,,
  \label{eq:actionmodel}
\end{equation}
where $\kappa_5$ is the five-dimensional gravitational constant, $\Lambda=-12/R^2$ is the cosmological constant, with $R$ being the AdS radius and $\hat{g}_\text{YM}$ the Yang-Mills coupling.
\\[\baselineskip]
The Einstein and Yang-Mills equations derived from the above action are
\begin{equation}
  \begin{aligned}
    R_{\mu\nu}+\frac{4}{R^2}g_{\mu\nu}&=\kappa_5^2\left(T_{\mu\nu}
      -\frac{1}{3}T^\rho_\rho g_{\mu\nu}\right)\,,\\
    \nabla_\mu F^{a\mu\nu}&=-\epsilon^{abc}A^b_\mu F^{c\mu\nu}\,,
  \end{aligned}
  \label{eq:Einstein-Yang-Mills}
\end{equation}
where the Yang-Mills energy-momentum tensor $T_{\mu\nu}$ is
\begin{equation}
  T_{\mu\nu}=\frac{1}{\hat{g}_\text{YM}^2}\left[F^a_{\mu\rho}{F^a_\nu}^\rho
    -\frac{1}{4}g_{\mu\nu}F^a_{\sigma\rho}F^{a\sigma\rho}\right]\,.
  \label{eq:TYangMills} 
\end{equation}
\\[\baselineskip]
A known solution of these equations of motion is the AdS Reissner-Nordström
black hole discussed in \ref{ssec:Normal-Phase-S-Wave} where
$\alpha=\kappa_5/\hat g_{\text{YM}}$. The gauge field is given by
\begin{equation}
  A=\mu\left(1-\frac{u^2}{u_{\mathrm{H}}^2}\right)\tau^3\dd t\,.
  \label{eq:RNgaugefieldpwave}
\end{equation}
where $\tau^3=\sigma^3/2\ci$ with $\sigma^3$ the third Pauli matrix. We consider the diagonal representations of the gauge group since we may rotate the flavor coordinates until the chemical potential lies in the third isospin direction.
\\[\baselineskip]
Another solution which corresponds to the holographic p-wave solution may be
obtained if we choose a gauge field ansatz \cite{Gubser:2008wv,Ammon:2009xh}
\begin{equation}
  A=\phi(r)\tau^3\dd t+w(r)\tau^1\dd x\,.\\
  \label{eq:gaugefieldansatz}
\end{equation}
The motivation for this ansatz is as follows: In the field theory we introduce
an isospin chemical potential by the the boundary values of the time components
of the gauge field $\phi$. This breaks the $SU(2)$ symmetry down to a diagonal
$U(1)$ which is generated by $\tau^3$. We denote this $U(1)$ as $U(1)_3$. In
order to study the transition to the superfluid state, we allow solutions with
non-zero $\langle J_x^1\rangle$ such that we include the dual gauge field
$A_x^1=w$ in the gauge field ansatz. Since we consider only isotropic and
time-independent solutions in the field theory, the gauge fields exclusively
depend on the radial coordinate $r$. With this ansatz the Yang-Mills
energy-momentum tensor defined in \eqref{eq:TYangMills} is diagonal. Solutions
with $\langle J_x^1\rangle\not=0$ also break the spatial rotational symmetry
$SO(3)$ down to $SO(2)$\footnote{Note that the finite temperature and chemical
  potential already break the Lorentz group down to $SO(3)$.} such that our
metric ansatz will respect only $SO(2)$. In addition, the system is invariant
under the $\mathds Z_2$ parity transformation $P_\parallel: x\to -x$ and $w\to
-w$. Furthermore, given that the Yang-Mills energy-momentum tensor is diagonal,
a diagonal metric is consistent,
\begin{equation}
  \dd s^2=-N(r)\sigma(r)^2\dd t^2+\frac{1}{N(r)}\dd r^2+r^2f(r)^{-4}
  \dd x^2+r^2f(r)^2\left(\dd y^2+\dd z^2\right)\,,
  \label{eq:metricansatz}
\end{equation}
with $N(r)=-2m(r)/r^2+r^2/R^2$.
\\[\baselineskip]
Inserting our ansatz into the Einstein and Yang-Mills equations leads to six
equations of motion for $m(r), \sigma(r),f(r),\phi(r),w(r),\psi(r)$ and one
constraint equation from the $rr$ component of the Einstein equations. The
dynamical equations may be written as
\begin{equation}
  \begin{aligned}
    m'&=\frac{\alpha^2 rf^4w^2\phi^2}{6N\sigma^2}
        +\frac{r^3\alpha^2\phi'^2}{6\sigma^2}
        +N\left(\frac{r^3f'^2}{f^2}+\frac{\alpha^2}{6}rf^4w'^2\right)\,,\\
    \sigma'&=\frac{\alpha^2f^4w^2\phi^2}{3rN^2\sigma}
             +\sigma\left(\frac{2rf'^2}{f^2}
             +\frac{\alpha^2f^4w'^2}{3r}\right)\,,\\
    f''&=-\frac{\alpha^2f^5w^2\phi^2}{3r^2N^2\sigma^2}
         +\frac{\alpha^2f^5w'^2}{3r^2}-f'\left(\frac{3}{r}-\frac{f'}{f}
         +\frac{N'}{N}+\frac{\sigma'}{\sigma}\right)\,,\\
    \phi''&=\frac{f^4w^2\phi}{r^2N}-\phi'\left(\frac{3}{r}
            -\frac{\sigma'}{\sigma}\right)\,,\\
    w''&=-\frac{w\phi^2}{N^2\sigma^2}
         -w'\left(\frac{1}{r}+\frac{4f'}{f}+\frac{N'}{N}
           +\frac{\sigma'}{\sigma}\right)\,.
  \end{aligned}
  \label{eq:eomsbr}
\end{equation}
The equations of motion are invariant under five scaling transformations
(invariant quantities are not shown),
\begin{equation}
  \begin{alignedat}{5}
    &\text{(I)} &\qquad \sigma&\to\lambda\sigma, &\qquad \phi&\to\lambda\phi,
    &\qquad&&\qquad&\\
    &\text{(II)} & f&\to\lambda f, & w&\to\lambda^{-2}w, &&&&\\
    &\text{(III)} & r&\to\lambda r, & m&\to\lambda^4m, & w&\to\lambda w,
    &\phi&\to\lambda\phi,\\
    &\text{(IV)} & r&\to\lambda r, & m&\to\lambda^2 m, &
    R&\to\lambda R, 
    &\phi&\to\lambda^{-1}\phi,\quad\;\alpha\to\lambda\alpha,
  \end{alignedat}
\label{eq:scaltrans}
\end{equation}
where in each case $\lambda$ is some real positive number. We use (I) and (II) to set the boundary values of both $\sigma$ and $f$ to one, so that the metric will be asymptotically $AdS$. Also we can use (III) to set $r_h$ to one, but we will keep it as a bookkeeping device. We use (IV) to set the AdS radius $R$ to one.

\subsection{Phase Diagram\label{ssec:Phase-Diagram}}
\begin{figure}[t]
  \centering
  \includegraphics[width=.5\textwidth]{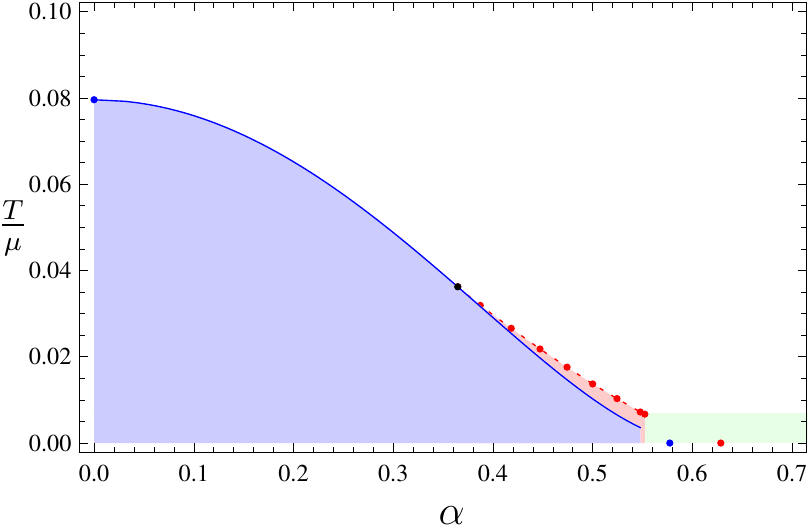}
  \caption[Phase Structure of Hairy Black Hole at Zero Baryon Chemical
  Potential]{Phase structure of the p-wave superconductor theory: In the blue 
    and red region the broken phase is the thermodynamically preferred phase, 
    while in the white region the Reissner-Nordström black hole is the ground 
    state. In the blue region the Reissner-Nordström black hole is unstable 
    and the transition from the white to the blue region is second order. In 
    the red region the Reissner-Nordström black hole is still stable. The 
    transition form the white to the red region is first order. The black dot 
    determines the critical point where the order of the phase transition 
    changes. In the green region we cannot trust our numerics.
    \label{fig:phasediag}}
\end{figure}
Let us now discuss the phase structure of this theory. At temperature below the critical temperature the thermodynamically favored phase is the holographic superfluid. By varying the parameter $\alpha$, the critical temperature changes. In addition in \cite{Ammon:2009xh} it was found that the order of the phase transition depends on the ratio of the coupling constants $\alpha$. For $\alpha\le \alpha_c=0.365$, the phase transition is second order while for larger values of $\alpha$ the transition becomes first order. The critical temperature decreases as we increase the parameter $\alpha$. The quantitative dependence of the critical temperature on the parameter $\alpha$ is given in Figure~\ref{fig:phasediag}. The broken phase is thermodynamically preferred in the blue and red region while in the white region the Reissner-Nordström black hole is favored. The Reissner-Nordström black hole is unstable in the blue region and the phase transition from the white to the blue region is second order. In the red region, the Reissner-Nordström black hole is still stable however the state with non-zero condensate is preferred. The transition from the white to the red region is first order. In the green region we cannot trust our numerics. At zero temperature, the data is obtained as described in \cite{Basu:2009vv,Gubser:2010dm}.

\subsection{Diffusion Constants\label{ssec:Diffusion-Constants}}
\begin{figure}[t]
  \centerline{
    \includegraphics[width=.48\textwidth]{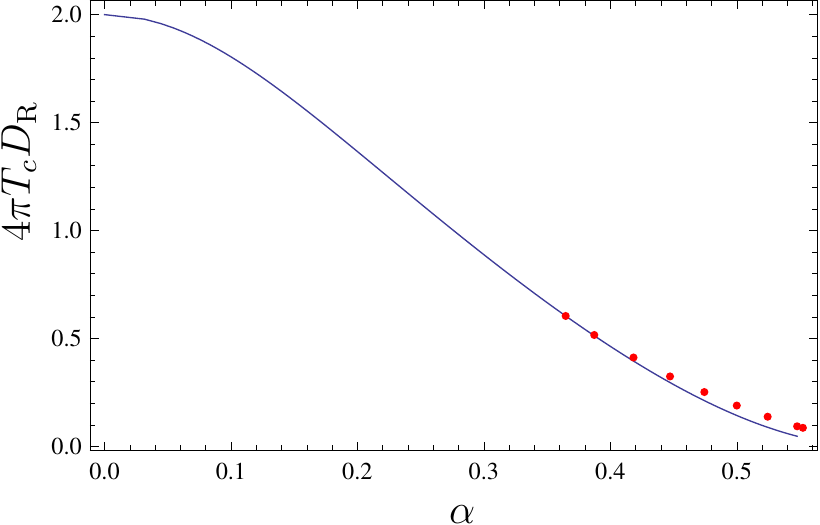}
    \hspace{.04\textwidth}
    \includegraphics[width=.48\textwidth]{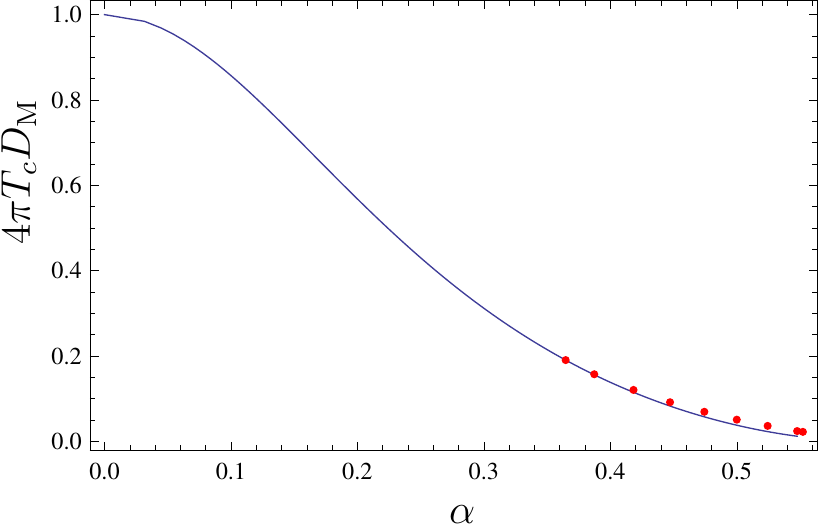}
  }
  \caption{$C_R = 4 \pi T_c D_R(T_c) $ and $C_M = 4 \pi T_c D_M (T_c)$ 
    related to charge diffusion $D_{\mathrm{R}}$ and
    momentum diffusion $D_{\mathrm{M}}$ depending on the strength of the
    backreaction $\alpha$ for the holographic p-wave superfluid with $d=4$. The
    blue line corresponds to the constants evaluated at the temperature at 
    which the Reissner-Nordström black hole becomes unstable (see blue line in
    Figure~\ref{fig:phasediag}). The red dots corresponds to the constants
    evaluated at the critical temperature at which the superfluid phase is
    thermodynamically preferred (see red dots in Figure~\ref{fig:phasediag}).
    \label{fig:DiffusionConstPwave}}
\end{figure}

We now determine the diffusion constants at the critical
temperature. Since the system is still described by a
Reissner-Nordström black hole as in the holographic s-wave
superfluid and thus the equations of motion for the fluctuations
coincide, we can use the same expressions to calculate the diffusion
constants for the holographic p-wave superfluid. The only change is the
dependence of the critical temperature on the parameter $\alpha$. The numerical
results are shown in Figure~\ref{fig:DiffusionConstPwave}. Comparing the phase
diagram of \figref{fig:phasediag} and the diffusion constants of
\figref{fig:DiffusionConstPwave}, we see that they are virtually identical. 
Furthermore, the analytical expression converting the curve of critical values
of $\nicefrac T\mu$ into the product $D(T_c)T_c$ is the same for both holographic superconductors. Thus, it is not surprising that we will find
answers that have a very large resemblance. Moreover, the same is true for the
s-wave superconductors and the comparison between the s-wave and the p-wave
superconductor: All curves for $D(T_c)T_c$ are very similar in both cases. In
the following section we will give a detailed conclusion and explain some
possible mechanisms which may lead to the decrease of $D(T_c) T_c$  in the
backreacted case.
%

%%% Local Variables: 
%%% mode: latex
%%% TeX-master: "homes-law"
%%% End: 

\section{Conclusions\label{sec:Conclusions}}

Let us summarize the main results of our paper: Motivated by Homes' law we have
analyzed the diffusion constants in holographic s- and p-wave superconductors
with backreaction. In particular we have discussed the temperature and density
dependence of the momentum and R-charge diffusion for various masses of the
scalar field (in the s-wave case) and for different strengths of the
backreaction. We have found that the diffusion constants decrease with
increasing backreaction and that the decay of the diffusion constants increases
with increasing mass of the scalar field. For non-negative masses of the s-wave
superconductor scalar field, we find an emerging $\text{AdS}_2$ at zero
temperature which defines a critical strength of the backreaction
$\alpha_c$. Above this critical backreaction there is no phase transition to the
superfluid/condensed phase, thus implying that $\alpha_c$ defines a quantum
phase transition between the normal and the condensed phase. These results are
of intrinsic interest within holography.
\\[\baselineskip]
Let us now discuss these results in relation to Homes' law: We have found that
without backreaction, holographic superconductors obey Homes' law since
$\tau_cT_c=\text{const.}$ in all spacetime dimensions greater than two. In the
case of s-wave superconductors, this holds for all scalar field masses. Without
backreaction, the result $\tau_cT_c=\text{const.}$ is almost automatic since the
diffusion constant $D$ scales as $1/T$. Turning on the backreaction, we find
that $D(T_c)T_c$ is no longer constant, but decreases with increasing
backreaction strength. Both for s-wave and p-wave holographic superconductors,
we obtain virtually the same curves for the various diffusion constants in 
three and four dimensions. In particular, the s-wave superconductor shows this
behavior in three and four dimensions for different masses of the scalar
field. This is a strong indication that there is a universal principle at work,
although we do not find a true constant. There are several possibilities how
corrections may arise:
\begin{enumerate}
\item The simplest explanation is the fact that we cannot assume the
  validity of the assumption \eqref{eq:Sum-Rule-Plasma-Frequency} for 
  holographic superconductors. As discussed at the end of Section
  \ref{ssec:Discussion}, increasing the backreaction $\alpha$ leads to the
  formation of a pseudo-gap and there are normal state charge carriers present
  in the superconducting phase as well. The dependence of the pseudo-gap on the
  backreaction has been studied holographically in \cite{Pan:2011ns} where it
  was found that the pseudo-gap arising in the superconducting phase becomes
  larger with increasing backreaction, as clearly visible in Figure 4 of 
  \cite{Pan:2011ns}. Assuming that the Ferrell-Glover-Tinkham sum rule
  \eqref{eq:Superconducting-Strength} is still valid in the presence of the
  pseudo-gap, $N_{\text s}$ in \eqref{eq:Superconducting-Strength} is no longer
  zero. This leads to correction terms to \eqref{eq:Sum-Rule-Plasma-Frequency}
  and \eqref{eq:Homes-Law-Sum-Rule3},
  \begin{alignat}{3}
    \omega_{\text{Ps}}^2&=\omega_{\text{Pn}}^2-8N_{\text s}
    &\qquad &\Leftrightarrow & \qquad 
    \tau_cT_c&=\frac{4\pi}C\left(1-\frac{N_{\text s}}{N_{\text n}}\right).
    \label{eq:Homes-Law-Correction}    
  \end{alignat}
  Thus, we would expect the ``constant'' in \eqref{eq:Homes-Law-Tanner3} to
  decrease and this is exactly what we see in \figref{fig:C-Constants-3D},
  \ref{fig:CM-Constants-4D}, \ref{fig:CR-Constants-4D} and 
  \ref{fig:DiffusionConstPwave}.
\item Even more dramatically, the sum rule \eqref{eq:Superconducting-Strength}
  may not be valid for holographic superconductors in the presence of the
  backreaction. In this case Homes' law will be implied by Tanner's law 
  \eqref{eq:Tanners-Law} which applies to high $T_c$ superconductors. It would
  be interesting to study Tanner's law in a holographic context (regardless of
  its relevance to Homes' law). Our result leads to the conjecture that the
  relation between the superconducting charge carrier and the normal state
  charge carrier concentration is dependent on the strength of the backreaction
  \begin{align}
    n_{\text s}&=B(\alpha)n_{\text n},
    \label{eq:Relation-Charge-Carrier}
    \intertext{such that \eqref{eq:Homes-Law-Tanner3} is modified by the 
      monotonically decreasing function $B(\alpha)$ yielding}
    \tau_cT_c&=4\pi\frac{B(\alpha)}C.
    \label{eq:Modified-Pseudo-Gap}
  \end{align}
\item The proportionality between the time scale and the diffusion constant 
  could in principle depend on $\alpha$ and $\bar Q$, i.e.~a function
  $A(\alpha)$, say. In this case the question arises if there might be some
  additional dynamics concerning diffusion in holographic
  superfluids/superconductors which would have to be taken into account. This
  would lead to the modified version of \eqref{eq:Homes-Law-Diffusion},
  \begin{equation}
    D(T_c)T_c=\frac{4\pi}{A(\alpha)C}.
    \label{eq:Modification-Diffusion}
  \end{equation}
\item From a condensed matter point of view, we should look at the dominant
  time scale which is given by the energy relaxation time. This needs
  not  necessarily to be the same as the momentum time scale. However, the 
  Einstein-Higgs models obey relativistic symmetries and therefore the
  momentum diffusion/time scales are related to the energy 
  diffusion/time scales. In fact, the energy diffusion constant is identical to
  the momentum diffusion as expected from the symmetries, up to a factor 
  depending on the spacetime dimensions: Conformal symmetry implies that the
  bulk viscosity vanishes, which ensures that hydrodynamic modes 
  corresponding to energy and momentum diffusion are related.\footnote{A 
    natural candidate for energy diffusion is the sound attenuation $\Gamma$ 
    following from the dispersion relation
    $\omega(k)=v_{\text{sound}}k-\ci\Gamma k^2$. For a conformal fluid,
    $\Gamma$ is determined by the momentum diffusion via 
    $\Gamma=\frac{d-2}{d-1}D_{\text M}$.} So the natural choice is indeed to
  look at momentum and R-charge diffusive processes to determine the relevant
  time scales, as done in this paper. Nevertheless, this issue deserves further
  investigation.
\item A more speculative and interesting correction could originate from the
  fact that we already have an infinite Drude peak in the normal phase 
  indicating an ideal conductor, 
  i.e.~$\re\sigma(\omega)\propto\delta(\omega)$, since there is no lattice
  present in our model. Entering the superconducting phase seems to add
  additional spectral weight due to the condensation of the scalar field. This
  in turn implies that the perfect conducting metal in the normal state
  \textit{does not} turn into a superconductor but that the ideal conductor and
  the superconductor coexist in the superconducting phase. Here we would again 
  expect a violation of equal spectral weight, or of the 
  Ferrell-Glover-Tinkham sum rule \eqref{eq:Superconducting-Strength},
  as well as having a simple relation between the normal state charge carrier
  concentration and the one in the superconducting state. In order to really
  understand what happens in the superconducting phase, one needs to determine 
  the behavior of $\rho_{\text s}$ for small temperatures and to check that only
  the contribution to the superconducting state is considered. It is believed
  that exactly at $T=0$ the normal state delta shaped Drude peak vanishes and
  the coefficient of the delta distribution is really the superfluid strength,
  yet this is not explicitly shown so far to our knowledge.
  \label{enum:Coexistence}
\end{enumerate}

\section{Outlook\label{sec:Outlook}}

As is explained in Section \ref{sec:Conclusions}, the most striking explanation
for corrections to $\tau_cT_c=\text{const.}$ in the backreacting case 
comes from additional degrees of freedom in the pseudo-gap. Let us give a list
of open questions which needs to be addressed to make further progress:
\begin{itemize}
\item An explicit calculation to verify the validity of the sum rules of
  holographic superconductors is in itself an interesting question. Therefore
  one needs to calculate the superfluid density including its dependence on the
  backreaction at zero temperature as well as the optical conductivity above 
  the critical temperature and near zero temperature. The conceptual setting is
  quite clear, but the numerical implementation might be challenging.
\item In order to confirm the assumptions made in \eqref{eq:Tanners-Law} one
  needs to check if the holographic superconductors show a relation between
  $n_{\text s}$ and $n_{\text n}$, i.e.~Tanner's law. The same challenges might
  arise as in the aforementioned point.
\item It would be interesting to understand if the holographic
  superconductors consists of an ideal metal coexisting with a superconductor
  following the speculative point \ref{enum:Coexistence} made in the
  conclusion. If so, one needs to devise a scheme to calculate the superfluid
  strength by removing the influence of the perfectly conducting metal. There
  are several ways to achieve this:  First of all it would be interesting to
  look for the Meißner-Ochsenfeld effect, i.e.~to calculate the transversal
  response of the gauge field which is only sensitive to the superfluid 
  strength deep in the superconducting phase. Alternatively, a further idea is
  to change the geometry such that the degrees of freedom of the normal state
  become very massive and are thus ``gapped out'' of the spectrum. Possible
  candidates are the hard-wall geometry, studied in the condensed matter 
  context in \cite{Sachdev:2011ze}, or a more smooth realization emerging in 
  the AdS-soliton solutions \cite{Nishioka:2009zj}, focusing in particular on 
  the AdS-soliton to AdS-soliton superconductor transition. A suppression of 
  normal low energy spectral weight has also been observed recently in a 
  different context in \cite{Hartnoll:2012wm}. It will also be interesting to 
  consider holographic systems with lattice structure, such as
  \cite{Horowitz:2012ky} and \cite{Liu:2012tr}, in the context explained
  here. These do not have the $\delta$-peak in the normal phase, as expected 
  for a system with an underlying lattice.
\item Since within condensed matter physics, energy diffusion is relevant to
  Homes' law, a systematic investigation of the relation between energy 
  diffusion and momentum diffusion in the holographic context appears to be 
  highly desirable, in particular in the presence of finite density and 
  backreaction. As explained in the conclusion, our assumption that energy and 
  momentum diffusion are related within holography is well-motivated. 
  Nevertheless this is a crucial issue which requires systematic study.
\item It would also be useful to calculate the dependence of fermionic
  excitations on the backreaction, for instance by generalizing the work of
  \cite{Gubser:2010dm} and \cite{Ammon:2010pg}.
\end{itemize}
In any case it is very important to disentangle the different contributions to
the $\delta$-peak in the superconducting phase in order to learn the 
differences which arise in holographic super\-fluids/super\-con\-duct\-ors as
compared to ``real world'' systems.

\acknowledgments

We like to thank Jan Zaanen for inspiring discussions, for sharing 
his invaluable insight into the phenomenology of superconductors and last but
not least for the hospitality of the Instituut-Lorentz where the final phase 
of the paper was completed. Moreover we also like to thank Martin Ammon and 
Andy O'Bannon for collaboration at an early stage of this project and for useful discussions. We also would like to thank Subir Sachdev, Koenraad
Schalm and Jonathan Shock for discussions.

%%% Local Variables: 
%%% mode: latex
%%% TeX-master: "homes-law"
%%% End: 

\appendix
\section*{Appendices}

\section{Plasma-Frequency\label{sec:Plasma-Frequency}}

Here we give a brief overview over the plasma frequency and the relations to 
the dielectric function and its role in terms of the superfluid density. 
Formally, the plasma frequency is defined as
\begin{equation}
  \frac{\omega_{\text P}^2}8=\int_0^\infty\dd\omega\re\sigma(\omega),
  \label{eq:Plasma-Definition}
\end{equation}
using the optical sum rule. Using the relations between the complex
dielectric function $\epsilon(\omega)$ and the complex optical conductivity
$\sigma(\omega)$,
\begin{alignat}{3}
  \epsilon&=\epsilon_\infty+\frac{4\pi\ci}\omega\sigma(\omega) & \qquad
  &\Rightarrow & \qquad 
  \re\sigma(\omega)&=\frac\omega{4\pi}\im\epsilon(\omega),
  \label{eq:Relation-Dielectric-Conductivity}
\end{alignat}
we may rewrite the plasma frequency in terms of the dielectric function as
\begin{equation}
  \omega_{\text P}^2=\frac2\pi\int_0^\infty\dd\omega\omega\im\epsilon(\omega).
  \label{eq:Plasma-Dielectric}
\end{equation}
With the help of the Kramers-Kronig relation for non-negative frequencies
\begin{align}
  \re f(\omega)&=\frac2\pi\mathcal P\int_0^\infty\dd\omega'
  \frac{\omega'\im f(\omega')}{\omega'^2-\omega^2}, &
  \im f(\omega)&=-\frac{2\omega}\pi\mathcal P\int_0^\infty\dd\omega'
  \frac{\re f(\omega')}{\omega'^2-\omega^2},
  \label{eq:Kramers-Kronig}
\end{align}
we may relate $\im\epsilon(\omega)$ to $\re\epsilon(\omega)-1$. Note that for
the dielectric function the polarization $P(\omega)$ is the response to the
applied electric field $E(\omega)$,
\begin{equation}
  4\pi P(\omega)=\left(\epsilon(\omega)-1\right)E(\omega).
  \label{eq:Polarization}
\end{equation}
Therefore a convenient form to write the Kramers-Kronig relation for the
dielectric function is to include the additional $-1$ in the real part of
$\epsilon(\omega)$,
\begin{align}
  \re\epsilon(\omega)-1
  &=\frac2\pi\mathcal P\int_0^\infty\dd\omega'
  \frac{\omega'\im\epsilon(\omega')}{\omega'^2-\omega^2}
  =-\frac2\pi\frac1{\omega^2}\mathcal P\int_0^\infty\dd\omega'
  \frac{\omega'\im\epsilon(\omega')}{1-\frac{\omega'^2}{\omega^2}}\notag\mnewl
  &\hspace{-7pt}\overset{\omega\gg\omega'}=-\frac2\pi\frac1{\omega^2}
  \int_0^\infty\dd\omega'\omega'\im\epsilon(\omega')
  =-\frac{\omega_{\text P}^2}{\omega^2},
  \label{eq:Kramers-Kronig-Dielectric}
\end{align}
where we are interested in the high frequency regime 
($\omega\gg\nicefrac1\tau$) since the sum rule is strictly valid only for
$\omega\to\infty$. In experiments we deal with finite frequencies only, and 
thus it is possible to extract the plasma frequency by the following
extrapolation of experimental data,
\begin{equation}
  \omega_{\text P}^2=\lim_{\omega\to0}
  \left(-\omega^2\re\epsilon(\omega)\right).
  \label{eq:Plasma-Dielectric2}
\end{equation}
As explained in the main text the superconducting plasma frequency determines
the frequency above which the superconductors becomes ``transparent'' in 
analogy with the normal metal plasma frequency. The reason for this terminology
follows from the fact that photons can only penetrate the superconductor for
length scales smaller than the London penetration depth $\lambda_{\text L}$
which corresponds to $\omega_{\text{Ps}}$. Here, the superconducting plasma
frequency should be understood with the aforementioned analogy to normal metals
in mind, described by the Drude-Sommerfeld form of the optical conductivity
\eqref{eq:Drude-Sigma}. In this case the (superconducting) plasma frequency is
given by
\begin{equation}
  \omega_{\text{Ps}}^2=8\frac{ne^2}m
  \int_0^\infty\dd{(\omega\tau)}\frac1{1+(\omega\tau)^2}
  =8\frac{ne^2}m\arctan(\omega\tau)\Big|_0^\infty
  =4\pi\frac{ne^2}m=\lambda_{\text L}^{-2}.
  \label{eq:Superconducting-Sum-Rule}
\end{equation} 
Experimentally, we cannot reach infinite frequencies and thus the sum rule is
modified by a cut-off frequency $\omega_c$, sometimes called the partial optical
sum rule of the Drude-Sommerfeld form
\begin{equation}
  \int_0^{\omega_{\text c}}\dd{(\omega\tau)}\frac1{1+(\omega\tau)^2}
  =8\frac{ne^2}m\arctan(\omega\tau)\Big|_0^{\omega_{\text c}}
  =\frac{\omega_{\text P}^2}{4\pi}\arctan(\omega_{\text c}\tau).
  \label{eq:Partial-Sum-Rule}
\end{equation} 
Expanding the inverse tangent function for $\omega_c\to\infty$, we get a series
expansion in the relaxation rate $\nicefrac1\tau$ which reads
\begin{align}
  \arctan\left(\omega_{\text c}\tau\right)&=\frac\pi 2-\frac1{\omega_{\text c}\tau}
  +\mathcal O\left(\frac1{\omega_{\text c}^3\tau^3}\right),
  \label{eq:Arctan-Expansion}
  \intertext{and thus}
  \int_0^{\omega_{\text c}}\dd\omega\re\sigma(\omega)
  &\approx\frac{\omega_{\text P}^2}8\left(1-\frac2\pi
    \frac1{\omega_{\text c}\tau}\right),
  \label{eq:Partial-Sum-Rule2}
\end{align}
from which  we recover \eqref{eq:Plasma-Definition} for
$\omega_{\text c}\to\infty$.

\section{Drude-Sommerfeld Model\label{sec:Drude-Model}}

In the previous sections we made use of some facts of Drude-Sommerfeld
theory, which we now discuss in some detail.
The original Drude model is a purely classical model describing a gas of
electrons diffusing through a fixed lattice of ions. These generate a 
positively charge background to balance the negative charge of the electron
gas. Even after the discovery of quantum mechanics, the Drude model can still 
be used but some modifications are needed: The electrons are described by a
fermionic gas obeying Fermi-Dirac statistics. This extension of the original
model is usually called Drude-Sommerfeld model. There are several ways to
arrive at the Drude formula used in \eqref{eq:Plasma-Direct-Current} each with
their own set of approximations. Essentially, there are three approaches:
\begin{enumerate}
\item A classical approach assuming the existence of an average relaxation time
  describing the relaxation to equilibrium after turning off an external
  electric field, i.e.~the relaxation time approximation which operationally
  replaces the collision integral by a linearized approximation including only
  the relaxation time scale and the Maxwell-Boltzmann distribution in the
  kinetic/Boltzmann transport equation. Strictly speaking, this derivation is
  borrowed from hydrodynamics where the collisions refer to a gas of weakly 
  interacting particles.
\item A semi-classical approach taking into account the quantum nature of the
  electron gas. Thus the free mean path is determined by the Fermi velocity of
  the electrons $\ell=v_{\text F}\tau$ and only electrons near the Fermi surface
  can participate.
\item A full microscopic approach considering an interacting electron gas
  employing Fermi liquid theory in combination with a diagrammatic
  expansion (and additional approximations such as the random phase 
  approximation). Here the assumptions of Fermi liquid theory, in
  particular  adiabatic continuity, are  needed. Furthermore, it is useful to
  introduce an effective frequency dependent mass $m^*(\omega)$ for the
  quasi-particles of Fermi liquid theory incorporating the effects of
  electron-phonon and electron-electron interactions as well as a frequency
  dependent effective time scale $\tau^*(\omega)$. However, in the case of
  impurity scattering the ratio $\nicefrac{\tau^*}{m^*}$ is identical to the
  bare values since the renormalization of the lifetime cancels the
  renormalization of the mass. This can be explicitly seen in a diagrammatic
  calculation of the current-current correlator using the Kubo formula. Here 
  the mass renormalization cancels since the diagrams included in the
  self-energy are also included in the two particle diagrams. This is in
  agreement with Fermi liquid theory stating that the current of the quasi
  particles is independent of the interaction.
  \label{enum:Full-Derivation}
\end{enumerate}
Since we do not know in general if there are really quasi-particles involved 
in strongly correlated systems (apart from heavy Fermi liquids), it seems to be
a good strategy not attempt to explain Homes' law microscopically, but to take
the ``effective field theory'' philosophy that focuses on the macroscopic 
rather than the microscopic degrees of freedom. In particular dealing with
``Planckian dissipation'' there is no particle transport and we are primarily
interested in quantum critical transport at finite density. Therefore we may
follow the standard derivation using the Kubo formula without resorting to the
aforementioned technicalities. A nice exposition of this derivation can be 
found in \cite{Dressel:2003ME}. Here we will just state the main ideas and the
result. We start with an interaction Hamiltonian
\begin{equation}
  \mathcal H_{\text{int}}=-\frac1c\vvec J\scp\vvec A,
  \label{eq:Interaction-Hamiltonian}
\end{equation}
and use the Kubo formula for the conductivity,
\begin{equation}
  \sigma(\omega,\vvec k)=\sum_s\frac1{\hbar\omega}\int\dd t
  \bracket s{\vvec J(0,\vvec k)\scp\cc{\vvec J(t,\vvec k)}}s\e[-\ci\omega t].
  \label{eq:Kubo-Interaction-Hamiltonian}
\end{equation}
The current density operator $\vvec J$ can be rewritten in terms of the 
momentum operator. Assuming an exponential decay with a single time scale for all current-current correlators (relaxation time approximation) and the dipole
approximation (the external electric field is constant over the characteristic
length scale which is surely valid for $v_{\text F}\ll c$) we arrive at
\begin{equation}
  \sigma(\omega)=\frac{e^2\tau}{m^*}\frac f{1-\ci\omega\tau},
  \label{eq:Optical-Conductivity-F}
\end{equation}
where $f$ describes the oscillator strength measuring the transition probability between two states given by
\begin{equation}
  f=2\sum_{j,s,s'}\frac{\abs{\bracket s{\vvec p_j}{s'}}^2}
          {m^*\hbar(\omega_s-\omega_{s'})}.
  \label{eq:Oscillator-Strength}
\end{equation}
Furthermore the oscillator strength can be evaluated under the assumption of
free electrons i.e.
\begin{alignat}{3}
  m^*\hbar(\omega_s-\omega_{s'})&=\frac{\hbar^2\vvec k^2}2 & \qquad &\text{and} 
  & \qquad \sum_{s,s'}\abs{\bracket s{\vvec p_j}{s'}}^2&=\expect{\vvec p_j^2}
  =\frac{\hbar^2\vvec k^2}4,
  \label{eq:Free-Particle-FSS}
\end{alignat}
since the self-interaction or electron-lattice interactions are already taken
care of in the renormalization of the mass $m^*(\omega)$ and the relaxation rate
$\tau^{-1}(\omega)=\tau^{-1}_{\text{impurity}}+\tau^{-1}_{\text{el-ph}}
+\tau^{-1}_{\text{el-el}}$. Inserting \eqref{eq:Free-Particle-FSS} into
\eqref{eq:Oscillator-Strength} the oscillator strength $f$ is given by the electron (or charge carrier) density $n$ and thus
\begin{equation}
  \sigma(\omega)=\frac{ne^2\tau}{m^*}\frac1{1-\ci\omega\tau}
  =\frac{\omega_{\text p}^2}{4\pi}\frac1{\frac1\tau-\ci\omega},
  \label{eq:Drude-Sigma}
\end{equation}
using the definition of the plasma frequency
\eqref{eq:Superconducting-Plasmafrequency} which in the limit of
$\omega\to0$ reduces to \eqref{eq:Plasma-Direct-Current}. If we assume the
correction of $\tau_{\text{el-ph}}$ and $\tau_{\text{el-el}}$ to be small
compared to $\tau_{\text{impurity}}$ we can follow the reasoning presented in point
\ref{enum:Full-Derivation} and replace $\nicefrac{\tau^*}{m^*}$ by
$\nicefrac\tau m$. The maximum of the real part of the optical conductivity
\eqref{eq:Drude-Sigma} at $\omega=0$ is called the Drude peak, whereas the
imaginary part shows a maximum at $\omega=\nicefrac1\tau$.

\section{Equations of Motion for the s--Wave Fluctuations \label{sec:EOM-Fluctuations}}

First we expand the matter Lagrangian up to quadratic order in the fluctuations
\begin{equation}
  \begin{aligned}
    \mathcal L^{(0)}_{cd}&=-\left(\nabla_c\Phi+\ci A_c\Phi\right)
                           \left(\nabla_d\Phi-\ci A_d\Phi\right)-m^2\Phi^2, 
                           \mnewl
    \mathcal L^{(1)}_{cd}&=-\nabla_c\delta\cc\phi\nabla_d\Phi
                                -\nabla_d\delta\phi\nabla_c\Phi
                                +\ci\nabla_c\Phi A_d\delta\phi
                                -\ci\nabla_d\Phi A_c\delta\cc\phi
                                -\ci A_c\nabla_d\delta\psi\Phi \mnewl
                           &\hspace{12pt}+\ci A_d\nabla_c\delta\cc\phi\Phi
                                 -A_cA_d\Phi\left(\delta\phi
                                 +\delta\cc\phi\right)
                                   -\left(A_ca_d+A_da_c\right)\Phi^2
                                -m^2\Phi\left(\delta\phi+\delta\cc\phi\right),
                             \mnewl
    \mathcal L^{(2)}_{cd}&=-\nabla_c\delta\phi\nabla_d\delta\cc\phi
                             -\ci A_c\nabla_d\delta\phi\delta\cc\phi
                             +\ci A_d\nabla_c\delta\cc\phi\delta\phi
                             -\ci a_c\nabla_d\delta\phi\Phi
                             +\ci a_d\nabla_c\delta\cc\phi\Phi \mnewl
                          &\hspace{12pt}+\ci\nabla_c\Phi a_d\delta\phi
                             -\ci\nabla_d\Phi a_c\delta\cc\phi
                             -A_cA_d\delta\phi\delta\cc\phi
                             -\left(A_ca_d+A_da_c\right)
                             (\delta\phi+\delta\cc\phi) \mnewl
                          &\hspace{12pt}-a_aa_b\Psi^2-m^2\delta\cc\phi
                             \delta\phi,
    \end{aligned}
  \label{eq:Lagrangian-Phi-CD}
\end{equation}
as well as the Maxwell Lagrangian
\begin{multline}
  \mathcal L_{\text M}^{(1)}=\alpha^2L^2\sqrt{-G}\left[-\frac14G^{ce}G^{df}
    \left(F_{cd}\delta F_{ef}+\delta F_{cd}F_{ef}\right)
    +G^{cd}\mathcal L^{(2)}_{cd}\right. \mnewl
    -\frac14\left(G^{cs}G^{et}h_{st}G^{df}+G^{ds}G^{ft}h_{st}G^{ce}
      +\frac12G^{st}h_{st}G^{ce}G^{df}\right)F_{cd}F_{ef} \mnewl
    \left.+\left(G^{cs}G^{dt}h_{st}+\frac12G^{st}h_{st}G^{cd}\right)
      \mathcal L^{(0)}_{cd}\right],
  \label{eq:Backreaction-Fluctuation-Matter-First-Order}
\end{multline}
\begin{multline}
  \mathcal L_{\text M}^{(2)}=\alpha^2L^2\sqrt{-G}\left\{-\frac14G^{ce}G^{df}
    \delta F_{cd}\delta F_{ef}+G^{cd}\mathcal L^{(2)}_{cd}\right. \mnewl
  -\frac14\left(G^{cs}G^{et}h_{st}G^{df}+G^{ds}G^{ft}h_{st}G^{ce}
    +\frac12G^{st}h_{st}G^{ce}G^{df}\right)\left(F_{cd}\delta F_{ef}
    +\delta F_{cd}F_{ef}\right) \mnewl
  +\left(G^{cs}G^{dt}h_{st}+\frac12G^{st}h_{st}G^{cd}\right)
  \mathcal L^{(1)}_{cd} \mnewl
  -\frac14\left[G^{cs}G^{et}h_{st}G^{dm}g^{fn}h_{mn}
    +\frac12G^{mn}h_{mn}\left(G^{cs}G^{et}h_{st}G^{df}+G^{ds}G^{ft}h_{st}g^{ce}\right)
  \right.\mnewl
  +\left.\left(\frac18g^{st}G^{mn}-\frac14G^{tm}G^{sn}\right)h_{st}h_{mn}
    G^{ce}G^{df}\right]F_{ce}F_{df} \mnewl
  +\left.\left[\frac12G^{st}h_{st}G^{cm}G^{dn}h_{mn}
      +\left(\frac18G^{st}G^{mn}h_{st}h_{mn}-\frac14G^{tm}G^{sn}\right)h_{st}h_{mn}
      G^{cd}\right]\mathcal L^{(0)}_{cd}\right\}.
  \label{eq:Backreaction-Fluctuation-Matter-Second-Order}
\end{multline}
The corresponding equations of motion for the scalar fluctuations are involved
expressions so we will simplify them according to our needs. Since we are
working in the normal phase we can set $\Phi(u)\equiv0$. Furthermore, the only
non-vanishing component of the background gauge field is $A_t(u)$ with all
other components being zero. In order to work out the quasi-normal modes we
apply a Fourier transformation and assume plain wave behavior for the spatial dependence 
\begin{equation}
  \begin{aligned}
    \delta\phi(t,\vvec x,u)&=\int\frac{\dd\omega\dd[\mathit d--1]
      {\vvec k}}{(2\pi)^d}
    \e[-\ci\omega t+\ci\vvec k\scp\vvec x]\delta\phi(u), \mnewl
    a_a(t,\vvec x,u)&=\int\frac{\dd\omega\dd[\mathit d--1]{\vvec k}}{(2\pi)^d}
    \e[-\ci\omega t+\ci\vvec k\scp\vvec x]a_a(u), \mnewl
    h_{ab}(t,\vvec x,u)&=\int\frac{\dd\omega\dd[\mathit d--1]{\vvec k}}{(2\pi)^d}
    \e[-\ci\omega t+\ci\vvec k\scp\vvec x]h_{ab}(u).
  \end{aligned}
  \label{eq:Fourier-Transformation-Fluctuations}
\end{equation}
Thus, we end up with the following equation of motion for the scalar fluctuations
\begin{equation}
  \delta\phi''(u)+\left(\frac{f'(u)}{f(u)}-\frac{d-1}u\right)\delta\phi'(u)
  +\left[\frac{(\omega+A_t)^2}{f(u)^2}-\frac{\vvec k^2}{f(u)}
    -\frac{L^2m^2}{u^2f(u)}\right]\delta\phi(u)=0,
  \label{eq:Scalar-Fluctuations-Appendix}
\end{equation}
and for the gauge field fluctuations we have
\begin{multline}
  a_t''(u)-\frac{d-3}ua_t'(u)-\left(\frac{\vvec k^2}{f(u)}\right)a_t(u)
  -\frac\omega{f(u)}\vvec k\scp\vvec a(u)
   +\ci\omega\left(a_u'(u)-\frac{d-3}ua_u(u)\right) \mnewl
  -\frac{3u^2A_t'(u)}{2L^2f(u)}h_{tt}'(u)+\frac{u^2A_t'(u)}{2L^2}
  \sum_{i=1}^{d-1}h_{ii}'(u)+\frac{3u^2f(u)A_t'(u)}{2L^2}h_{uu}'(u) \mnewl
  +\frac u{L^2}A_t'(u)\sum_{i=1}^{d-1}h_{ii}(u)
  +\frac{\ci uA_t'(u)}{L^2}\sum_{i=1}^{d-1}h_{ui}(u)k_i
  +\frac32\frac{u^2f(u)}{L^2}\left(\frac{f'(u)}{f(u)}+\frac2u\right)
  A_t'(u)h_{uu}(u)=0,
  \label{eq:a_t-Equation}
\end{multline}
\begin{multline}
  \vvec a''(u)+\left(\frac{f'(u)}{f(u)}-\frac{d-3}u\right)
  \vvec a'(u)+\left(\frac{\omega^2}{f(u)^2}-\frac{\vvec k_\perp^2}{f(u)}
  \right)\vvec a(u)
  +\frac{\vvec k}{f(u)}\big(\vvec k_\perp\scp\vvec a(u)\big) \mnewl
  +\frac{\omega\vvec k}{f(u)^2}a_t(u)
  -\ci\vvec k\left[a'_u(u)+\left(\frac{f'(u)}{f(u)}
      -\frac{d-3}u\right)a_u(u)\right] \mnewl
  -\frac{u^2A_t'(u)}{L^2f(u)}\vvec h_t'(u)
  -\frac{2uA_t'(u)}{L^2f(u)}\vvec h_t(u)
  -\frac{\ci u^2\omega A_t'(u)}{L^2f(u)}\vvec h_u(u)=0,
  \label{eq:a_x-Equation}
\end{multline}
\begin{multline}
  \frac\omega{f(u)}a_t'(u)+\vvec k\scp\vvec a'(u)-\ci\left(\vvec k^2
    -\frac{\omega^2}{f(u)}\right)a_u(u) \mnewl
  -\frac32\frac{u^2\omega A_t'(u)}{L^2f(u)^2}h_{tt}(u)
  -\frac{u^2A_t'(u)}{L^2f(u)}\sum_{i=1}^{d-1}h_{ti}k_i
  +\frac{u^2\omega A_t'(u)}{2L^2f(u)}\sum_{i=1}^{d-1}h_{ii}(u)
  +\frac32\frac{u^2\omega A_t'(u)}{L^2}h_{uu}(u)=0.
  \label{eq:a_u-Equation}
\end{multline}
where $\vvec h_t(u),\vvec h_u(u)$ denotes the spatial entries of the
corresponding row/column and $\vvec k_\perp$ denotes the transverse vector
orthogonal to the directions given by the particular equations for the 
component of $\vvec a(u)$. For instance, we can look at the equation for one of
the $d-1$ spatial components, $\vvec a(u)=a_x(u)\vvec e_x$ say, so
$\vvec k_\perp=k_y\vvec e_y+k_z\vvec e_z+\dotsm$ and hence 
$\vvec k_\perp\scp\vvec a$ couples the complementary components to the one
chosen, i.e.~$a_y(u),a_z(u),\dotsc$. Furthermore, assuming $\vvec k=k_x\vvec
e_x$ , we see that the equations of the $a_x(u)$ component couples only with the
$a_t(u)$ component (in this case $\vvec k_\perp=\vvec 0$), whereas the equations
for all other $d-2$ spatial components decouple. Looking at the transverse
directions e.g. $\vvec k=k_y\vvec e_y+k_z\vvec e_z+\dotsm$, the equation for the
fluctuations in the $x$-direction will decouple from all other $d-2$, but the
equations for the remaining $d-2$ fluctuations will couple with each other and
$a_t(u)$.

%%% Local Variables: 
%%% mode: latex
%%% TeX-master: "homes-law"
%%% End: 

\bibliography{homes-law}

\end{document}